\documentclass[structabstract]{aa}

\usepackage{graphicx}
\usepackage[usenames]{color}

\begin{document}

\title{
Multiplicity of rapidly oscillating Ap stars\thanks{Based on
observations obtained at the European Southern Observatory,
Paranal, Chile (ESO programme Nos.~079.D-0537(A)
and 076.C-0170).}
}

   \author{M.~Sch\"oller\inst{1}
          \and
          S.~Correia\inst{2}
          \and
          S.~Hubrig\inst{2}
          \and
          D.~W.~Kurtz\inst{3}
          }

   \institute{European Southern Observatory,
              Karl-Schwarzschild-Str.~2,
              85748 Garching, Germany\\
              \email{mschoell@eso.org}
         \and
             Leibniz-Institut f\"ur Astrophysik Potsdam (AIP),
             An der Sternwarte 16,
             14482 Potsdam, Germany
          \and
             Jeremiah Horrocks Institute,
             University of Central Lancashire,
             Preston PR1 2HE, United Kingdom
             }

   \date{Received September 15, 1996; accepted March 16, 1997}

 
  \abstract
   {
Rapidly oscillating Ap (roAp) stars have rarely been found in binary or
higher order multiple systems. 
This might have implications for their origin.
   }
   {
We intend to study the multiplicity of this type of chemically peculiar stars, 
looking for visual companions in the range of angular separation between
0\farcs05 and 8\arcsec{}.
   }
   {
We carried out a survey of 28 roAp stars using diffraction-limited near-infrared
imaging with NAOS-CONICA at the VLT.
Additionally, we observed three non-oscillating magnetic Ap stars.
   }
   {
We detected a total of six companion candidates with low chance projection
probabilities.
Four of these are new detections, the other two are confirmations.
An additional 39 companion candidates are very likely chance projections.
We also found one binary system among the non-oscillating magnetic Ap stars.
The detected companion candidates have apparent K magnitudes between 6\fm8 and 19\fm5
and angular separations ranging from 0\farcs23 to 8\farcs9,
corresponding to linear projected separations of 30--2400\,AU.
   }
   {
While our study confirms that roAp stars are indeed not very often members of binary or multiple systems,
we have found four new companion candidates that are likely physical companions.
A confirmation of their status will help understanding the origin of the roAp stars.
   }


   \maketitle

\section{Introduction}
\label{sect:intro}



Rapidly oscillating Ap (roAp) stars are ideal targets for asteroseismology.
By comparing the observed frequency spectrum with the asymptotic 
pulsation theory, it is possible 
to specify their rotation period, their temperature, luminosity, radius, mass,
their atmospheric structure, their evolutionary status, and the geometry of their magnetic field.
More than forty roAp stars are known at present, with effective temperatures between 6400\,K and 8400\,K.  
Kurtz et al.\ (\cite{Kurtz2006}) list 35 roAp stars;
more can be found in individual papers since then.
They pulsate in high-overtone, low-degree, nonradial $p$-modes, with periods in the
range from 5.6 to 21\,min and typical amplitudes of a few millimagnitudes (e.g.\ Kurtz \cite{Kurtz1982}).

The roAp phenomenon is confined to a well-defined region of the Str\"omgren photometry 
parameter space (Martinez \cite{Martinez1993}).
However, this region also contains other Ap stars, in 
which no pulsation could be detected, despite thorough searches.
These apparently constant Ap stars (non-oscillating Ap stars, or noAp stars) appear remarkably similar to the 
roAp stars in many respects (color indices, abundances, magnetic fields).

A decade ago, 
after the completion of a study of the kinematical properties of rapidly 
oscillating Ap and noAp stars, Hubrig et al.\ (\cite{Hubrig2000})
realized that
none of the roAp stars is known to be a spectroscopic binary (SB).
They obtained several radial velocity measurements for a majority
of the roAp stars, but found no evidence for variations in any of these stars. 
The situation is quite different for other Ap stars co-existing in the 
same region of the H-R diagram.
In contrast with roAp stars, many noAp stars for which radial velocity data exist are known 
as spectroscopic binaries or show radial velocity variations.
The interpretation of this 
difference and its significance for the understanding of the origin of the pulsations 
in roAp stars is not clear so far.  
That until now no roAp star is known to be a spectroscopic binary is also in 
direct contrast to the situation for other types of pulsating variables 
(e.g., $\beta$\,Cep stars, $\delta$ Sct stars, or classical Cepheids), which are 
frequently found in SB systems. 

Neither theoretically nor observationally is our present knowledge 
sufficient to decide confidently whether tidal interaction in binaries may reduce the 
amplitude of or inhibit the pulsations in cool Ap stars. 
To establish this, a necessary 
condition would be to show that essentially all noAp stars are close binaries, or, alternatively,
to investigate whether all roAp stars are single stars or wide visual binaries.

Our search in the literature and catalogs for known double or 
multiple systems among roAp stars revealed that the three roAp stars $\gamma$\,Equ (HD\,201601),
$\beta$\,CrB, and $\alpha$\,Cir, which are the brightest and best studied stars 
in the group of roAp stars, do have optical or physical companions. 
Stelzer et al.\ (\cite{Stelzer2011}) combined data from the
Washington Double Star Catalog (Mason et al.\ \cite{Mason2001})
and from {\em Hipparcos} with the measurement reported in this paper
and derived a preliminary orbit for $\gamma$\,Equ.
They estimated the mass of the companion to be 0.6$\pm$0.4\,$M_{\sun}$.
The companion in the $\beta$\,CrB system at a separation of
0\farcs3 was frequently observed by speckle interferometry in the eighties. 
The companion in the system $\alpha$\,Cir is of spectral type K5V
(Eggleton \& Tokovinin \cite{EggletonTokovinin2008}).
All other roAp stars are considerably fainter than $\gamma$\,Equ, $\beta$\,CrB,
and $\alpha$\,Cir, and 
therefore have not been intensely studied with high resolution imaging instruments. 
Still, companions are reported for HD\,99563
and HR\,3831 (e.g.\ Dommanget \& Nys \cite{DommangetNys2002}).

The interpretation of a difference in duplicity between roAp and noAp stars and its meaning 
for the understanding of the origin of the pulsations in roAp stars is far from obvious. 
Even if the different internal structure of roAp stars is the reason for their pulsations,
it is difficult to understand why no roAp star was found to be in a close binary.

%

In the following we report the results of our multiplicity study
of this class of objects using NACO K-band imaging.

\section{Observations and data reduction}
\label{sect:observations}

We carried out observations of 28 roAp stars with NAOS-CONICA
(NACO; Lenzen et al.\ \cite{Lenzen2003}; Rousset et al.\ \cite{Rousset2003})
on the VLT in service mode between April and September 2007.
Furthermore, three ordinary Ap stars were observed in December 2005 and January 2006.
We used the S13 camera of CONICA, which provides the smallest available pixel scale of 13.3\,milliarcsec
and a field-of-view of 13\farcs6.
All data were collected through a $Ks$ filter in image autojitter mode, where the object is observed at typically
20 different image positions with random offsets between them.
Since all our sources are bright in $V$, we used the visible wavefront sensor of NAOS.

\begin{table*}
\centering
\caption{
Objects studied in the program.
}
\label{tab:objects}
\begin{tabular}{rccccr @{$\pm$} l}
\hline
\hline
\multicolumn{1}{c}{HD} &
\multicolumn{1}{c}{Other} &
\multicolumn{1}{c}{V} &
\multicolumn{1}{c}{K$^a$} &
\multicolumn{1}{c}{Spectral} &
\multicolumn{2}{c}{Parallax$^b$} \\
\multicolumn{1}{c}{number} &
\multicolumn{1}{c}{Identifier} &
\multicolumn{1}{c}{} &
\multicolumn{1}{c}{} &
\multicolumn{1}{c}{Type} &
\multicolumn{2}{c}{[mas]} \\
\hline
6532   & CD-27\,355   & 8.40  & 8.19 &  A3 Sr Cr      & 6.14  & 1.03 \\
9289   & BD-11\,286   & 9.40  & 8.87 &  A3 Sr Eu Cr   & \multicolumn{2}{c}{} \\
12932  & BD-19\,384   & 10.36 & 9.46 &  A4 Sr Eu      & \multicolumn{2}{c}{} \\
19918  & CD-82\,53    & 9.37  & 8.78 &  A5 Sr Eu Cr   & 4.07  & 0.82 \\ 
24712  & HR\,1217     & 5.99  & 5.26 &  A9 Sr Eu Cr   & 20.32 & 0.39 \\ 
42659  & BD-15\,1299  & 6.77  & 6.36 &  A3 Sr Cr Eu   & 7.60  & 0.51 \\ 
86181  & CPD-58\,1700 & 9.39  & 8.68 &  F0 Sr         & 3.49  & 1.05 \\ 
99563  & BD-08\,3173  & 8.2   & 7.74 &  F0 Sr         & 3.92  & 1.15 \\ 
101065 & CD-46\,7232  & 8.02  & 6.92 &  F3 Ho         & 8.93  & 0.87 \\ 
116114 & BD-17\,3829  & 7.03  & 6.35 &  F0 Sr Cr Eu   & 7.71  & 0.55 \\ 
119027 & CD-28\,10204 & 9.91  & 9.09 &  A3 Sr Eu      & \multicolumn{2}{c}{} \\
122970 & BD+06\,2827  & 8.30  & 7.31 &  F0 Cr Eu Sr   & 8.67  & 0.70 \\ 
134214 & BD-13\,4081  & 7.47  & 6.67 &  F2 Sr Eu Cr   & 9.74 & 0.64 \\ 
137949 & BD-16\,4093  & 6.68  & 6.25 &  F0 Sr Eu Cr   & 11.28 & 0.67 \\ 
150562 & CD-48\,11127 & 9.95  & 8.88 &  A5: Eu Si?    & \multicolumn{2}{c}{} \\
154708 & CD-57\,6753  & 8.76  & 7.95 &  A2 Sr Eu Cr   & 6.75  & 0.96 \\ 
161459 & CD-51\,11145 & 10.4  & 9.54 &  A2 Eu Sr Cr   & \multicolumn{2}{c}{} \\
166473 & CD-37\,12303 & 7.92  & 7.45 &  A5 Sr Eu Cr   & \multicolumn{2}{c}{} \\
176232 & HR\,7167     & 5.91  & 5.30 &  A6 Sr         & 12.76 & 0.29 \\ 
185256 & CD-30\,17252 & 9.92  & 8.92 &  F0 Sr Eu      & \multicolumn{2}{c}{} \\
190290 & CD-79\,800   & 9.89  & 9.18 &  A0 Eu Sr      & \multicolumn{2}{c}{} \\
193756 & CD-52\,9483  & 9.19  &      &  A9 Sr Cr Eu   & \multicolumn{2}{c}{} \\
196470 & BD-18\,5731  & 9.79  & 9.17 &  A2 Sr Eu      & \multicolumn{2}{c}{} \\
201601 & HR\,8097     & 4.71  & 4.01 &  A9 Sr Eu      & 27.55 & 0.62 \\ 
203932 & CD-30\,18600 & 8.82  & 8.13 &  A5 Sr Eu      & \multicolumn{2}{c}{} \\
213637 & BD-20\,6447  & 9.58  & 8.54 &  F1 Eu Sr      & \multicolumn{2}{c}{} \\
217522 & CD-45\,14901 & 7.53  & 6.63 &  A5 Sr Eu Cr   & 11.36 & 0.79 \\
218495 & CD-64\,1414  & 9.34  & 9.03 &  A2 Eu Sr      & \multicolumn{2}{c}{} \\
\hline
\multicolumn{7}{c}{Magnetic Ap stars} \\
\hline
40711  & BD+10 973 &    8.58 & 8.15 &  A0 Sr Cr Eu    & 2.66 & 0.95 \\
55719  & HR\,2727 &     5.31 & 5.14 &  A3 Sr Cr Eu    & 7.93 & 0.38 \\
59435  & V827\,Mon &    7.97 & 6.39 &  A4 Sr Cr Eu    & 2.70 & 0.76 \\
\hline
\end{tabular}  
\tablefoot{
In Col.~1, we give the HD number of the objects,
in Col.~2 another identifier, in Cols.~3 and 4 the magnitudes
in $V$ and $K$ bands,
in Col.~5 the spectral type,
and finally in Col.~6 the parallax.
Most information was collected from the SIMBAD database, except for the spectral
types, which are from Renson \& Manfroid (\cite{RensonManfroid2009}), and the parallaxes, obtained from
van Leeuwen (\cite{vanLeeuwen2007}).\\
$^a$For one target, SIMBAD does not list a K magnitude.\\
$^b$For some targets, Leeuwen (\cite{vanLeeuwen2007}) does not list a parallax.
}
\end{table*}


The sample was picked from the list of 35 roAp stars listed in Kurtz et al.\ (\cite{Kurtz2006})
for their accessibility from the VLT during the observing period.
$\alpha$\,Cir would also have been observable, but was well studied in the past
by other authors.
Since the presence of companions is relatively rare among the Ap stars,
we additionally chose 
three non-oscillating magnetic Ap stars because the presence of a companion
was mentioned in the literature 
(Renson \& Manfroid \cite{RensonManfroid2009} for HD\,40711 and HD\,59435;
Pourbaix et al.\ \cite{Pourbaix2004} and Mason et al.\ \cite{Mason2001} for HD\,55719).
The sample is listed in Table~\ref{tab:objects}.

The data reduction was performed with the eclipse package (Devillard \cite{Devillard1997}) in the standard way. 
Sky background frames obtained from median averaging
of the jittered frames were subtracted from the individual frames.
All frames were then flat-fielded and corrected for bad pixels
using calibration files provided by ESO.

After identifying all companion candidates in the resulting images visually,
we obtained astrometry and relative photometry of the 
multiple systems using the IRAF package DAOPHOT.
Since for some sources the errors determined by DAOPHOT were small, we adopted as a floor for
systematics 1/10$^{\rm th}$ of a pixel for the position of
the individual objects and 1\% for the flux ratio.
The final errors in the relative positions are estimated by combining quadratically the 
rms variations in our astrometric analysis with the 
uncertainty in the plate scale (13.26$\pm$0.03\,mas)
and detector orientation ($\pm$0.5$^{\circ}$), both provided by Masciadri et al.\ (\cite{Masciadri2003}).

\section{Results}
\label{sect:results}

\begin{table*}
\centering
\caption{
Astrometric and photometric results of the candidate binaries and multiples resolved in our study.
}
\label{tab:astrometry}
{\scriptsize
\begin{tabular}{r@{}lrr @{$\pm$} rr @{$\pm$} rr @{$\pm$} rr @{$\pm$} rr @{$\pm$} rr @{$\pm$} rr @{$\pm$} rc}
\hline
\hline
\multicolumn{2}{c}{HD} &
\multicolumn{1}{c}{MJD} &
\multicolumn{2}{c}{Separation} &
\multicolumn{2}{c}{Position} &
\multicolumn{2}{c}{K mag } &
\multicolumn{2}{c}{K mag } &
\multicolumn{2}{c}{K mag } &
\multicolumn{2}{c}{K mag } &
\multicolumn{2}{c}{Projected} &
\multicolumn{1}{c}{Chance} \\
\multicolumn{2}{c}{number} &
\multicolumn{1}{c}{} &
\multicolumn{2}{c}{} &
\multicolumn{2}{c}{angle} &
\multicolumn{2}{c}{difference} &
\multicolumn{2}{c}{system} &
\multicolumn{2}{c}{primary} &
\multicolumn{2}{c}{secondary} &
\multicolumn{2}{c}{linear} &
\multicolumn{1}{c}{projection } \\
\multicolumn{1}{c}{} &
\multicolumn{1}{c}{} &
\multicolumn{1}{c}{} &
\multicolumn{2}{c}{} &
\multicolumn{2}{c}{} &
\multicolumn{2}{c}{} &
\multicolumn{2}{c}{} &
\multicolumn{2}{c}{} &
\multicolumn{2}{c}{} &
\multicolumn{2}{c}{separation} &
\multicolumn{1}{c}{probability} \\
\multicolumn{1}{c}{} &
\multicolumn{1}{c}{} &
\multicolumn{1}{c}{} &
\multicolumn{2}{c}{[\arcsec{}]} &
\multicolumn{2}{c}{[$^{\circ}$]} &
\multicolumn{2}{c}{} &
\multicolumn{2}{c}{} &
\multicolumn{2}{c}{} &
\multicolumn{2}{c}{} &
\multicolumn{2}{c}{[AU]} &
\multicolumn{1}{c}{[\%]} \\
\hline
\multicolumn{18}{c}{Binaries}\\
\hline
           9289 & & 54311.35 & 0.441 & 0.003 &   72.7 & 0.7 &  1.70 & 0.01 &  8.87 & 0.02 &  9.08 & 0.03 & 10.77 & 0.05 & \multicolumn{2}{c}{} & 2.16$\times10^{-4}$ \\
          12932 & & 54311.37 & 0.239 & 0.003 &  171.4 & 0.9 &  0.65 & 0.01 &  9.46 & 0.02 &  9.94 & 0.03 & 10.59 & 0.06 & \multicolumn{2}{c}{} & 4.77$\times10^{-5}$ \\
 $^{\ast}$99563 & & 54210.04 & 1.784 & 0.004 &  216.6 & 0.5 &  0.66 & 0.01 &  7.74 & 0.02 &  8.21 & 0.03 &  8.87 & 0.05 & 455.1 & 133.5 & 1.77$\times10^{-3}$ \\
         101065 & & 54212.11 & 8.648 & 0.005 &  140.9 & 0.5 &  7.45 & 0.04 &  6.92 & 0.02 &  6.92 & 0.02 & 14.37 & 0.06 & 968.4 & 94.3 & 5.47 \\
         185256 & & 54233.35 & 4.965 & 0.005 &  306.7 & 0.5 &  8.61 & 0.06 &  8.92 & 0.03 &  8.92 & 0.03 & 17.53 & 0.08 & \multicolumn{2}{c}{} & 6.31 \\
         196470 & & 54262.37 & 6.921 & 0.003 &  190.9 & 0.5 &  6.73 & 0.02 &  9.17 & 0.02 &  9.17 & 0.02 & 15.90 & 0.04 & \multicolumn{2}{c}{} & 6.29 \\
$^{\ast}$201601 & & 54266.25 & 0.829 & 0.003 &  256.8 & 0.5 &  2.71 & 0.01 &  4.01 & 0.26 &  4.09 & 0.26 &  6.80 & 0.29 & 30.1 & 0.7 & 3.82$\times10^{-4}$ \\
         203932 & & 54262.40 & 0.227 & 0.003 &   98.2 & 0.9 &  1.10 & 0.01 &  8.13 & 0.02 &  8.46 & 0.03 &  9.57 & 0.06 & \multicolumn{2}{c}{} & 7.18$\times10^{-5}$ \\
\hline
\multicolumn{18}{c}{Systems with more than one companion candidate}\\
\hline
            86181&AB & 54223.07 & 6.560 & 0.004 &     116.9 & 0.5 &      6.04 & 0.01 &      8.68 & 0.02     &  8.68 & 0.02         & 14.71 & 0.03 & 1879.7 & 565.5 & 24.7 \\
            86181&AC &          & 5.549 & 0.003 &      94.7 & 0.5 &      6.80 & 0.02 & \multicolumn{2}{c}{} & \multicolumn{2}{c}{} & 15.48 & 0.04 & 1590.0 & 478.4 & 29.5 \\
            86181&AD &          & 3.532 & 0.004 &     294.3 & 0.5 &      8.43 & 0.04 & \multicolumn{2}{c}{} & \multicolumn{2}{c}{} & 17.11 & 0.06 & 1012.0 & 304.5 & 14.3 \\
            86181&AE &          & 7.309 & 0.004 &     137.6 & 0.5 &      8.97 & 0.05 & \multicolumn{2}{c}{} & \multicolumn{2}{c}{} & 17.65 & 0.07 & 2094.3 & 630.1 & 48.5 \\
            86181&AF &          & 8.222 & 0.005 &      44.3 & 0.5 &      9.42 & 0.07 & \multicolumn{2}{c}{} & \multicolumn{2}{c}{} & 18.10 & 0.09 & 2355.9 & 708.8 & 56.8 \\
            86181&AG &          & 6.848 & 0.005 &      47.7 & 0.5 &      9.66 & 0.08 & \multicolumn{2}{c}{} & \multicolumn{2}{c}{} & 18.33 & 0.09 & 1962.2 & 590.3 & 44.1 \\
            86181&AH &          & 3.068 & 0.006 &     267.5 & 0.5 &      9.94 & 0.09 & \multicolumn{2}{c}{} & \multicolumn{2}{c}{} & 18.61 & 0.11 & 879.1  & 264.5 & 11.0 \\
            86181&AI &          & 5.088 & 0.005 &     268.8 & 0.5 &      9.97 & 0.09 & \multicolumn{2}{c}{} & \multicolumn{2}{c}{} & 18.64 & 0.11 & 1457.9 & 438.6 & 27.5 \\
            86181&AJ &          & 4.256 & 0.009 &     342.1 & 0.5 &     10.22 & 0.10 & \multicolumn{2}{c}{} & \multicolumn{2}{c}{} & 18.90 & 0.12 & 1219.5 & 366.9 & 20.1 \\
            86181&AK &          & 6.497 & 0.005 &     293.0 & 0.5 &     10.29 & 0.11 & \multicolumn{2}{c}{} & \multicolumn{2}{c}{} & 18.97 & 0.12 & 1861.6 & 560.1 & 40.8 \\
            86181&AL &          & 5.560 & 0.007 &      34.0 & 0.5 &     10.40 & 0.11 & \multicolumn{2}{c}{} & \multicolumn{2}{c}{} & 19.08 & 0.13 & 1593.1 & 479.3 & 31.9 \\
\hline
           150562&AB & 54222.34 & 7.872 & 0.003 &     193.0 & 0.5 &      4.43 & 0.01 &      8.88 & 0.02     &  8.88 & 0.02         & 13.31 & 0.03 & \multicolumn{2}{c}{} & 45.8 \\
           150562&AC &          & 2.122 & 0.003 &       1.5 & 0.5 &      5.58 & 0.01 & \multicolumn{2}{c}{} & \multicolumn{2}{c}{} & 14.46 & 0.03 & \multicolumn{2}{c}{} & 9.06 \\
           150562&AD &          & 4.459 & 0.003 &     111.1 & 0.5 &      6.05 & 0.01 & \multicolumn{2}{c}{} & \multicolumn{2}{c}{} & 14.93 & 0.03 & \multicolumn{2}{c}{} & 37.3 \\
           150562&AE &          & 3.492 & 0.004 &     299.1 & 0.5 &      6.28 & 0.01 & \multicolumn{2}{c}{} & \multicolumn{2}{c}{} & 15.15 & 0.04 & \multicolumn{2}{c}{} & 25.1 \\
           150562&AF &          & 4.725 & 0.004 &     125.0 & 0.5 &      6.52 & 0.01 & \multicolumn{2}{c}{} & \multicolumn{2}{c}{} & 15.40 & 0.04 & \multicolumn{2}{c}{} & 41.3 \\
           150562&AG &          & 5.024 & 0.004 &     138.3 & 0.5 &      6.91 & 0.02 & \multicolumn{2}{c}{} & \multicolumn{2}{c}{} & 15.79 & 0.04 & \multicolumn{2}{c}{} & 45.5 \\
           150562&AH &          & 3.224 & 0.004 &     258.0 & 0.5 &      8.38 & 0.04 & \multicolumn{2}{c}{} & \multicolumn{2}{c}{} & 17.26 & 0.06 & \multicolumn{2}{c}{} & 22.2 \\
           150562&AI &          & 2.813 & 0.004 &     282.9 & 0.5 &      8.43 & 0.04 & \multicolumn{2}{c}{} & \multicolumn{2}{c}{} & 17.31 & 0.06 & \multicolumn{2}{c}{} & 17.4 \\
           150562&AJ &          & 1.139 & 0.005 &     101.8 & 0.6 &      8.52 & 0.04 & \multicolumn{2}{c}{} & \multicolumn{2}{c}{} & 17.40 & 0.06 & \multicolumn{2}{c}{} & 3.08 \\
           150562&AK &          & 2.153 & 0.004 &     130.9 & 0.5 &      8.54 & 0.04 & \multicolumn{2}{c}{} & \multicolumn{2}{c}{} & 17.41 & 0.06 & \multicolumn{2}{c}{} & 10.6 \\
           150562&AL &          & 1.565 & 0.005 &      99.6 & 0.5 &      8.77 & 0.05 & \multicolumn{2}{c}{} & \multicolumn{2}{c}{} & 17.65 & 0.07 & \multicolumn{2}{c}{} & 5.74 \\
           150562&AM &          & 2.479 & 0.005 &     182.2 & 0.5 &      8.90 & 0.05 & \multicolumn{2}{c}{} & \multicolumn{2}{c}{} & 17.78 & 0.07 & \multicolumn{2}{c}{} & 13.8 \\
           150562&AN &          & 2.677 & 0.008 &     128.8 & 0.5 &      9.94 & 0.09 & \multicolumn{2}{c}{} & \multicolumn{2}{c}{} & 18.82 & 0.11 & \multicolumn{2}{c}{} & 15.9 \\
           150562&AO &          & 2.668 & 0.009 &     144.8 & 0.5 &     10.10 & 0.09 & \multicolumn{2}{c}{} & \multicolumn{2}{c}{} & 18.98 & 0.12 & \multicolumn{2}{c}{} & 15.8 \\
           150562&AP &          & 2.776 & 0.007 &      32.4 & 0.5 &     10.16 & 0.10 & \multicolumn{2}{c}{} & \multicolumn{2}{c}{} & 19.03 & 0.12 & \multicolumn{2}{c}{} & 17.0 \\
           150562&AQ &          & 2.055 & 0.009 &     182.6 & 0.5 &     10.60 & 0.12 & \multicolumn{2}{c}{} & \multicolumn{2}{c}{} & 19.48 & 0.15 & \multicolumn{2}{c}{} & 9.68 \\
           150562&AR &          & 2.830 & 0.018 &     158.8 & 0.6 &     10.61 & 0.12 & \multicolumn{2}{c}{} & \multicolumn{2}{c}{} & 19.49 & 0.14 & \multicolumn{2}{c}{} & 17.6 \\
\hline
           154708&AB & 54222.31 & 8.919 & 0.003 &     113.8 & 0.5 &      4.65 & 0.01 &      7.95 & 0.03     &      7.97 & 0.03     & 12.62 & 0.04 & 1321.3 & 187.9 & 3.88 \\
           154708&AC &          & 0.782 & 0.004 &      53.0 & 0.6 &      4.78 & 0.01 & \multicolumn{2}{c}{} & \multicolumn{2}{c}{} & 12.75 & 0.05 & 115.9  &  16.5 & 3.43$\times10^{-2}$ \\
           154708&AD &          & 7.608 & 0.003 &      89.0 & 0.5 &      8.85 & 0.06 & \multicolumn{2}{c}{} & \multicolumn{2}{c}{} & 16.81 & 0.09 & 1127.1 & 160.3 & 38.0 \\
           154708&AE &          & 5.185 & 0.004 &     184.2 & 0.5 &      9.32 & 0.07 & \multicolumn{2}{c}{} & \multicolumn{2}{c}{} & 17.29 & 0.10 & 768.1  & 109.2 & 20.2 \\
           154708&AF &          & 9.246 & 0.008 &     342.1 & 0.5 &      9.59 & 0.08 & \multicolumn{2}{c}{} & \multicolumn{2}{c}{} & 17.56 & 0.11 & 1369.8 & 194.8 & 51.3 \\
\hline
           161459&AB & 54210.08 & 8.820 & 0.006 &      12.6 & 0.5 &      8.10 & 0.05 &      9.54 & 0.02     &      9.54 & 0.02     & 17.64 & 0.07 & \multicolumn{2}{c}{} & 50.6 \\
           161459&AC &          & 7.981 & 0.011 &     231.7 & 0.5 &      8.87 & 0.07 & \multicolumn{2}{c}{} & \multicolumn{2}{c}{} & 18.41 & 0.10 & \multicolumn{2}{c}{} & 43.9 \\
\hline
           193756&AB & 54258.37 & 8.835 & 0.005 &     144.3 & 0.5 &      7.37 & 0.04 &      8.65 & 0.02     &      8.65 & 0.02     & 16.02 & 0.06 & \multicolumn{2}{c}{} & 7.73 \\
           193756&AC &          & 6.045 & 0.010 &     132.9 & 0.5 &      8.53 & 0.07 & \multicolumn{2}{c}{} & \multicolumn{2}{c}{} & 17.17 & 0.09 & \multicolumn{2}{c}{} & 3.85 \\
\hline
\multicolumn{18}{c}{Magnetic Ap stars}\\
\hline
$^{\ast}$55719 & & 53729.34 & 0.714 & 0.003 & 265.0 & 0.6 & 2.97 & 0.01 & 5.14 & 0.02 & 5.21 & 0.02 & 8.18 & 0.04  & 90.0 & 4.3 & 5.67$\times10^{-4}$ \\
\hline
\end{tabular}  
}
\tablefoot{
In Col.~1, we list the HD number for each system as well as the pair designation for
the systems with more than one companion candidate, and
Col.~2 gives the modified Julian date for the observations.
In Cols.~3, 4, and 5, we show the separation, position angle (from North to East), and magnitude difference
in the K band between the components, as retrieved by aperture photometry from our images.
In Col.~6, we give the K band magnitude for the whole system, as derived from the 2MASS or
DENIS catalogs,
and in Cols.~7 and 8 we give K band magnitudes for the primary and secondary component, as
determined from Cols.~5 and 6.
Whenever we have a parallax available, we give the projected linear separation in Col.~9.
Column~10 finally lists the chance projection probability of the secondary component,
as described in Sect.~\ref{sect:projections}.
An asterisk preceding the HD number in Col.~1 indicates systems where the companion was
known before our study.
}
\end{table*}

We were able to detect companion candidates around 13 roAp stars and one magnetic Ap star.
The astrometric and photometric results are presented in Table~\ref{tab:astrometry} for all 
multiple systems of our sample. 

\begin{figure*}
\centering
\includegraphics[width=0.32\textwidth, angle=0]{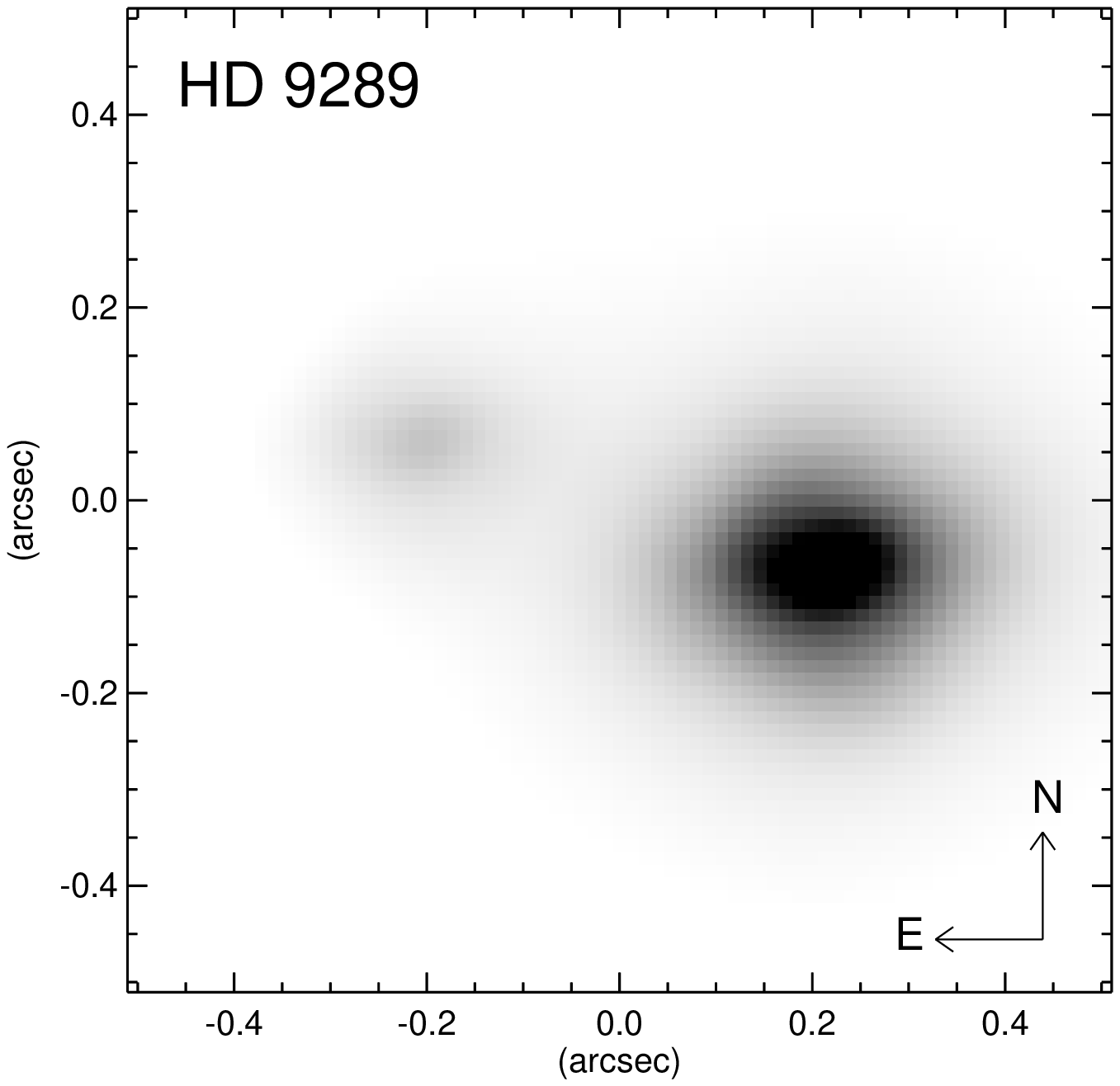}
\includegraphics[width=0.32\textwidth, angle=0]{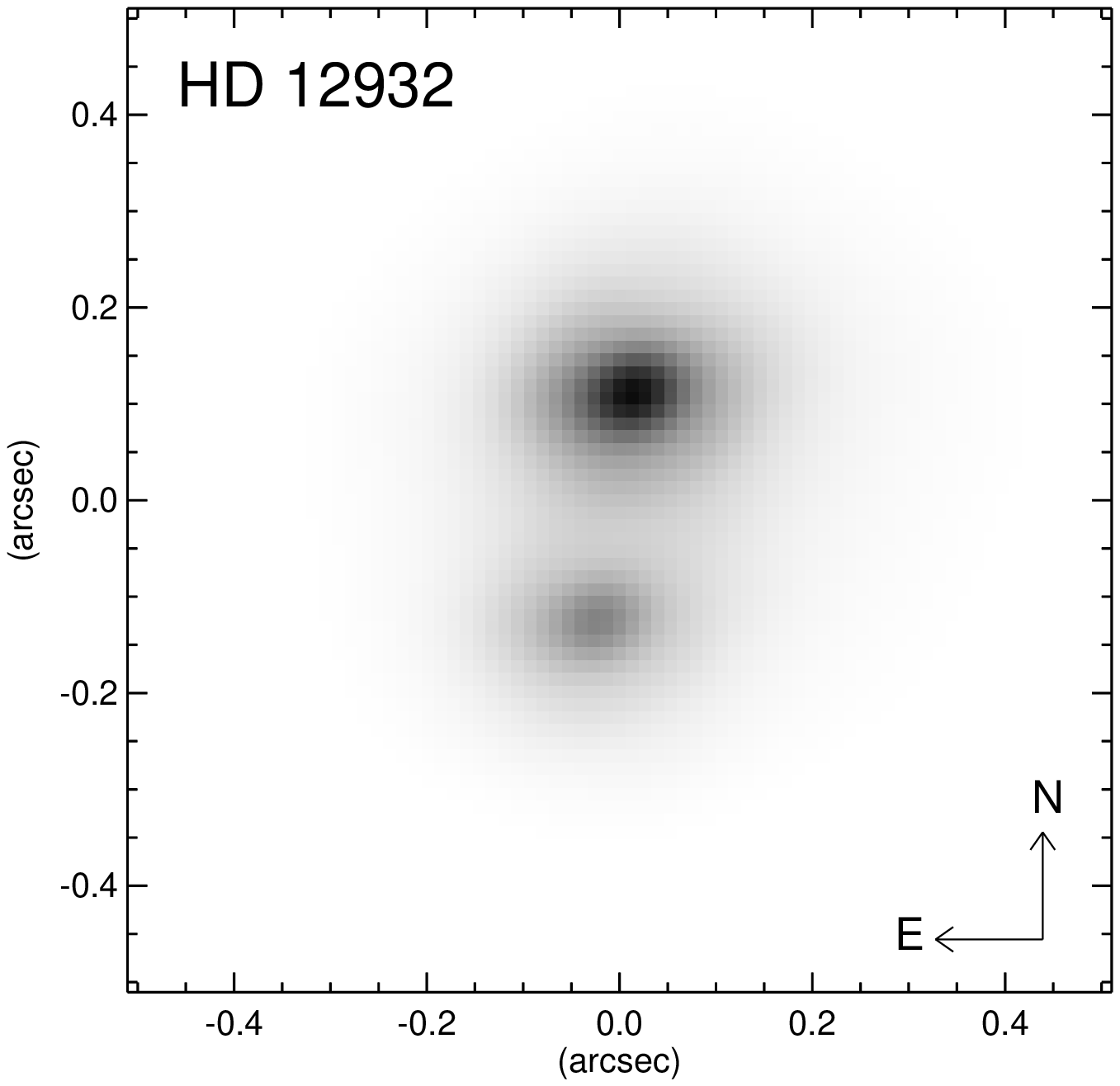}
\includegraphics[width=0.32\textwidth, angle=0]{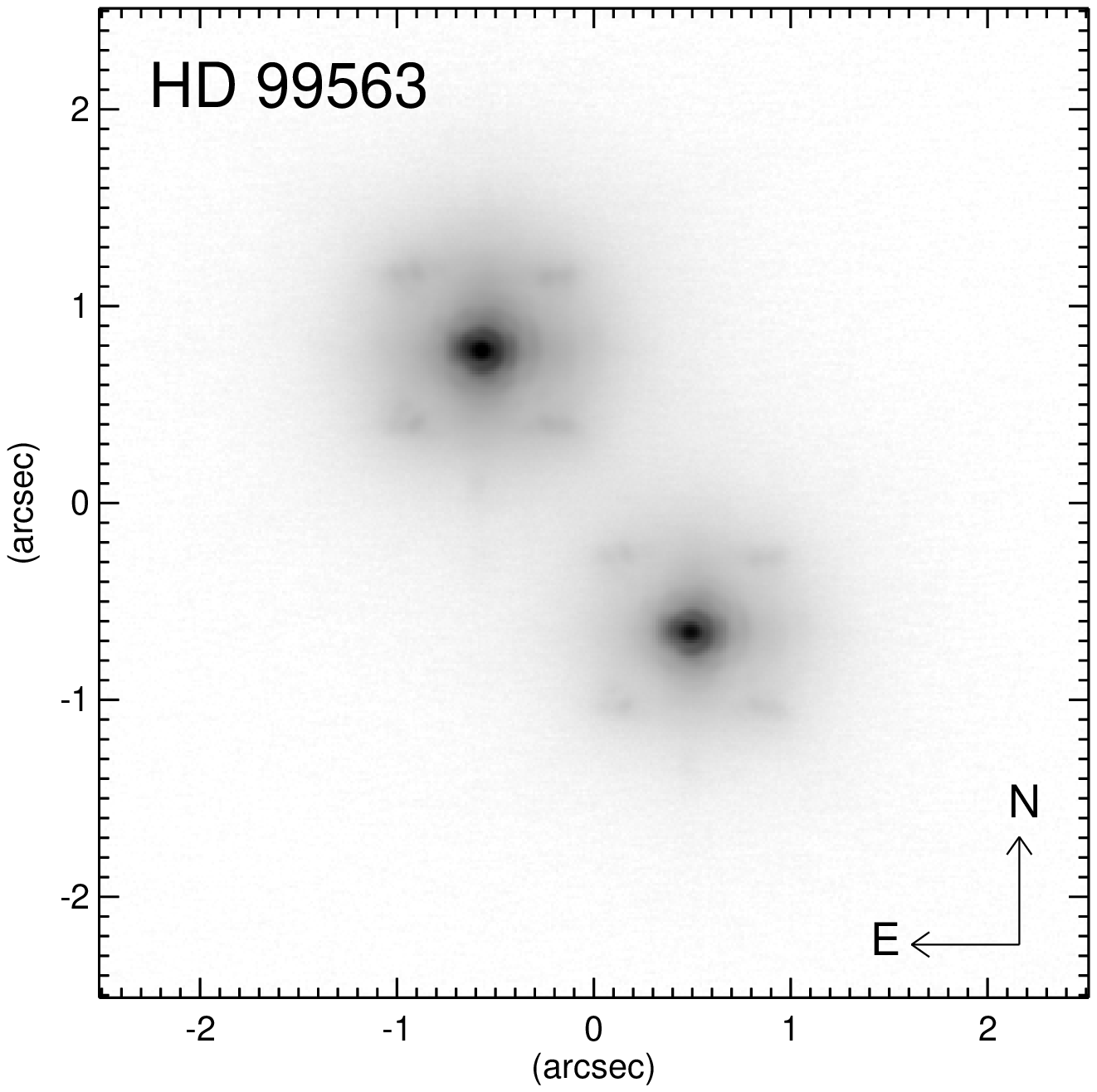}
\includegraphics[width=0.32\textwidth, angle=0]{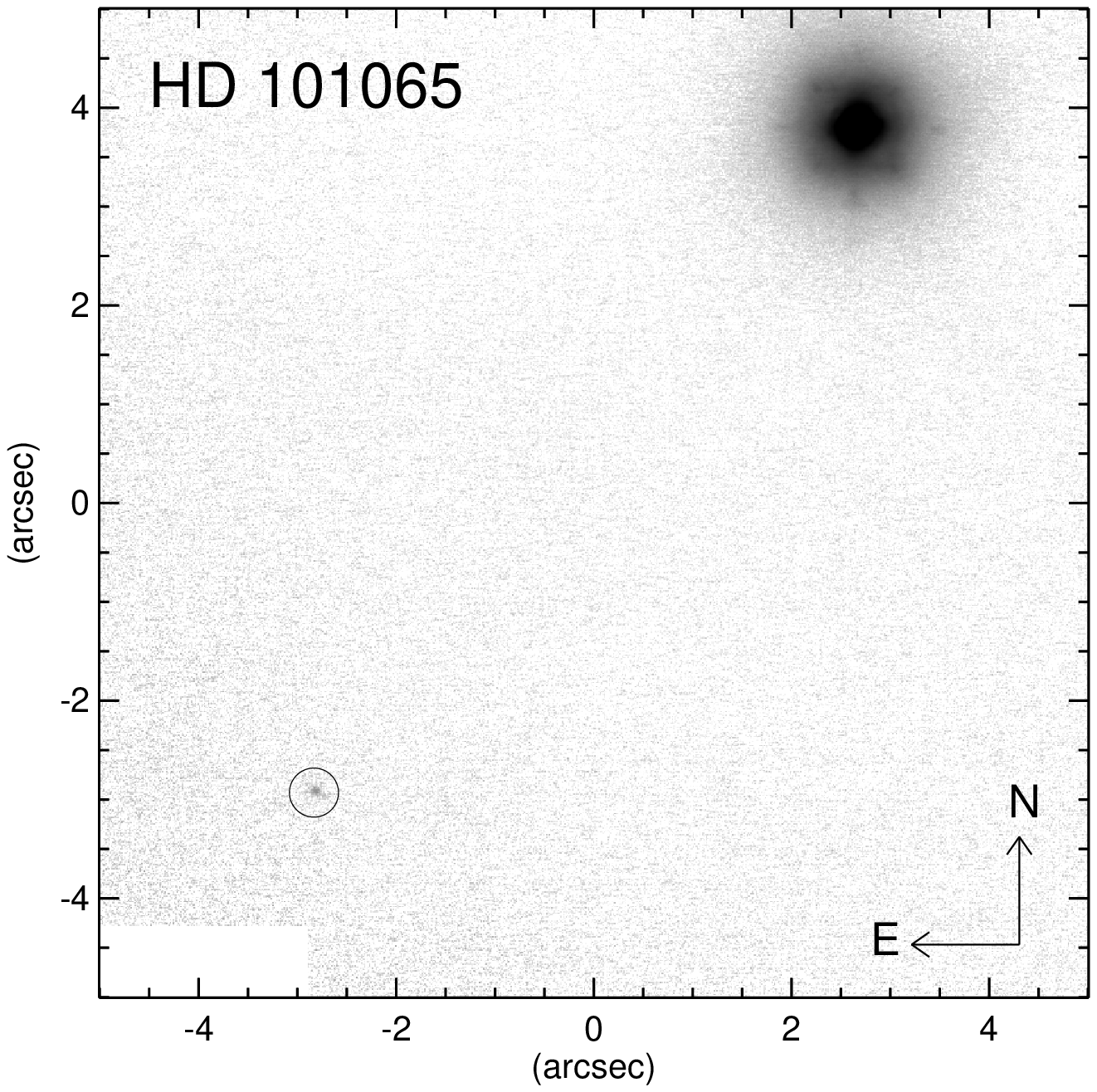}
\includegraphics[width=0.32\textwidth, angle=0]{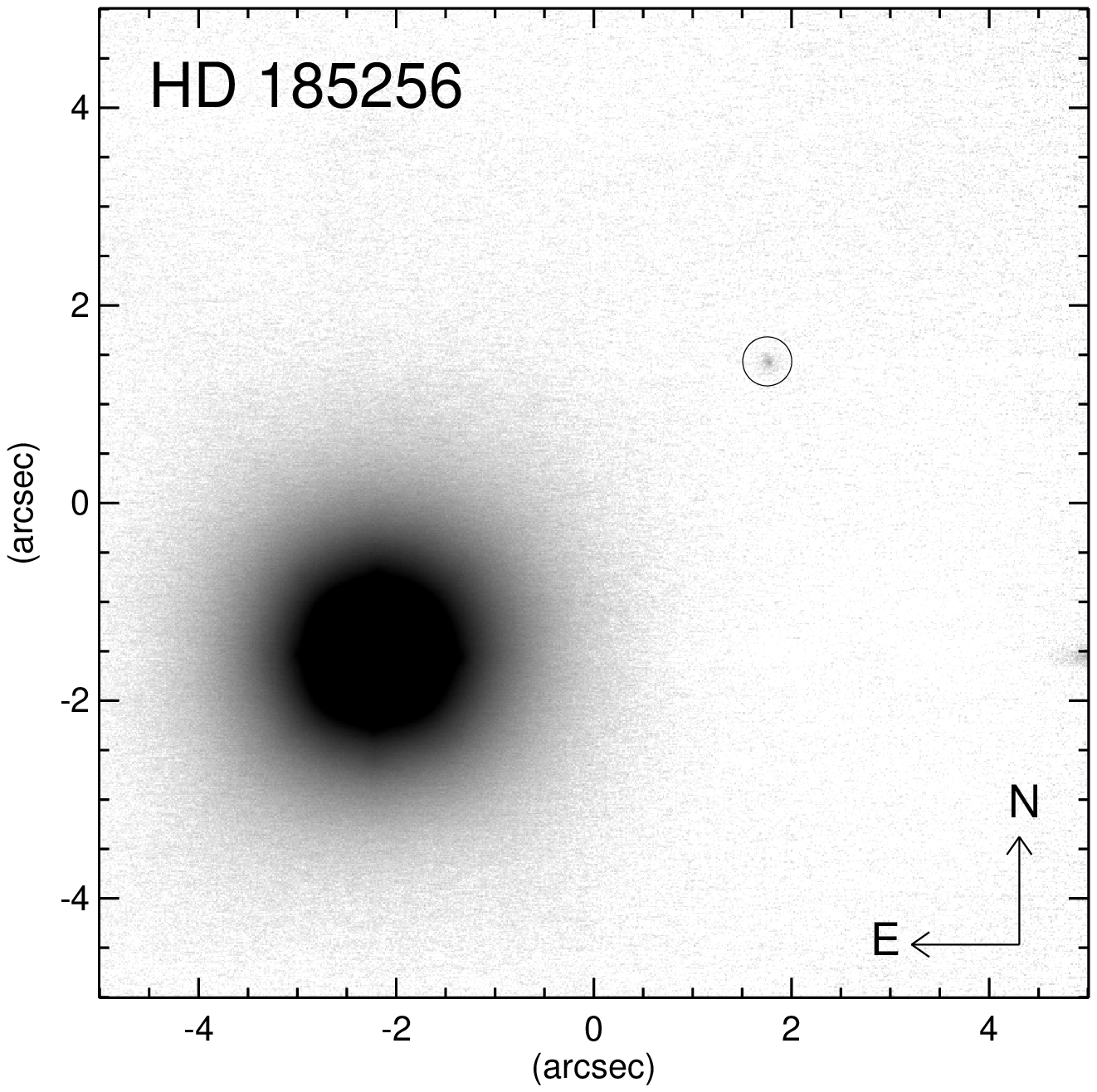}
\includegraphics[width=0.32\textwidth, angle=0]{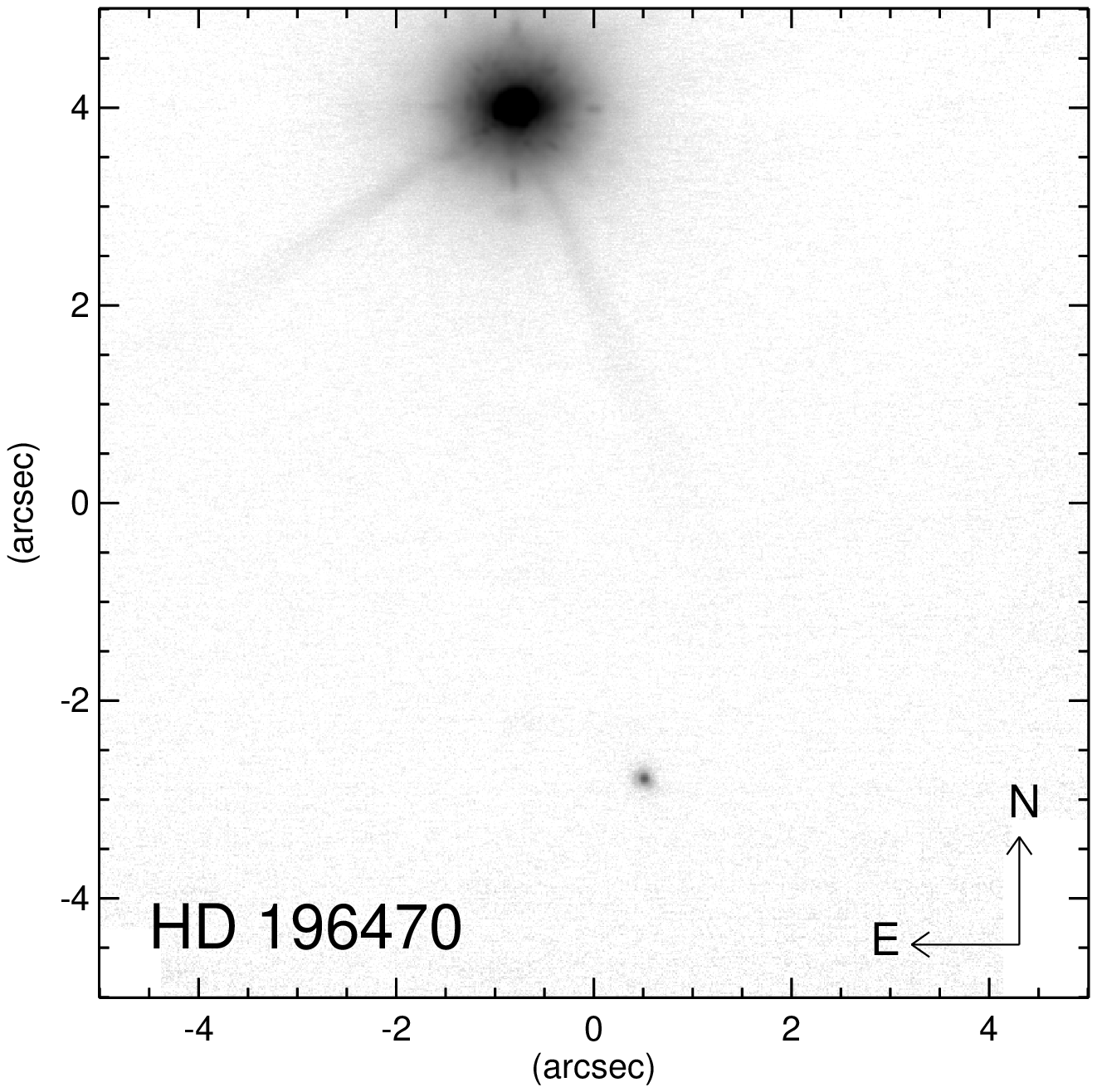}
\includegraphics[width=0.32\textwidth, angle=0]{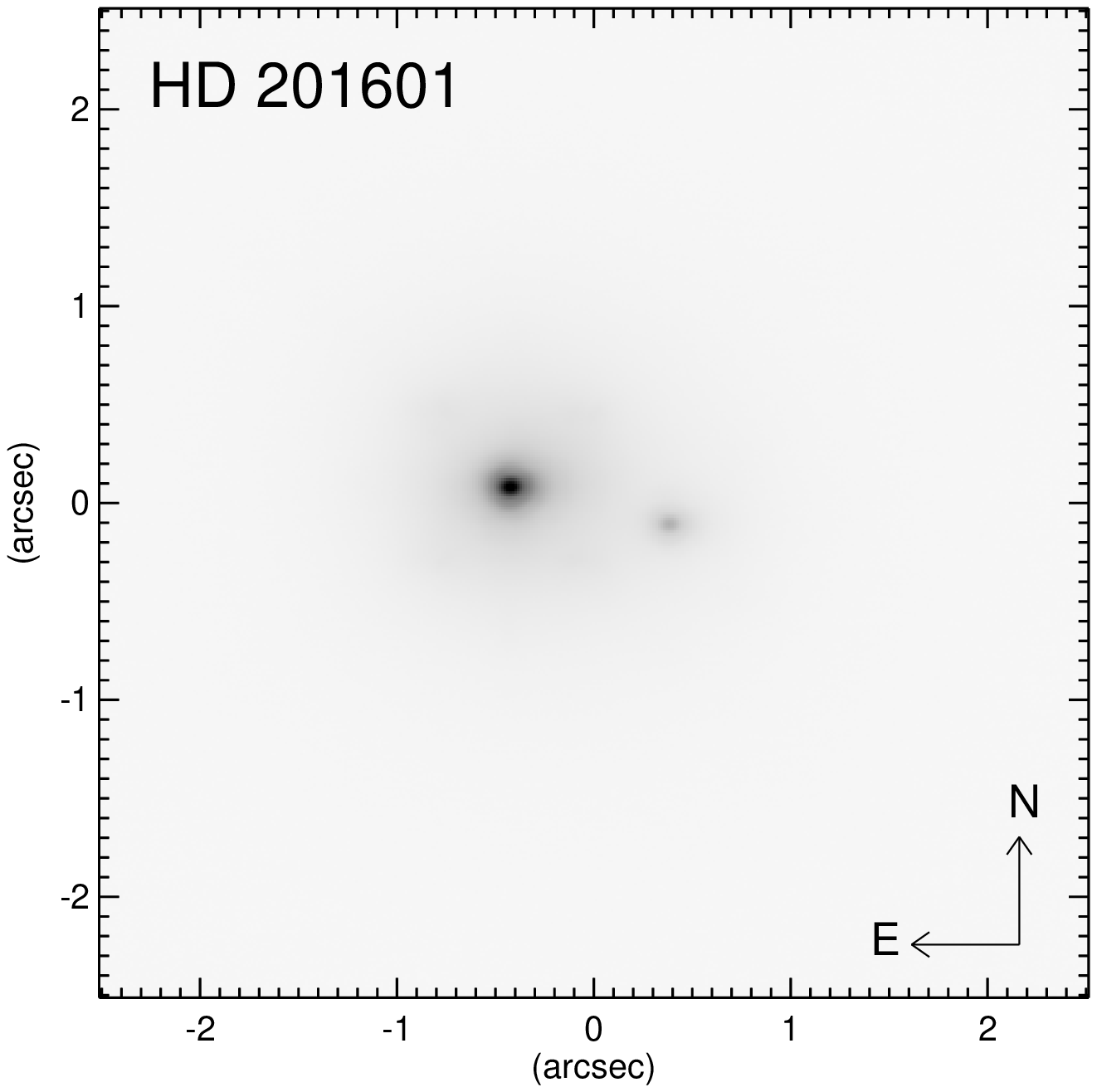}
\includegraphics[width=0.32\textwidth, angle=0]{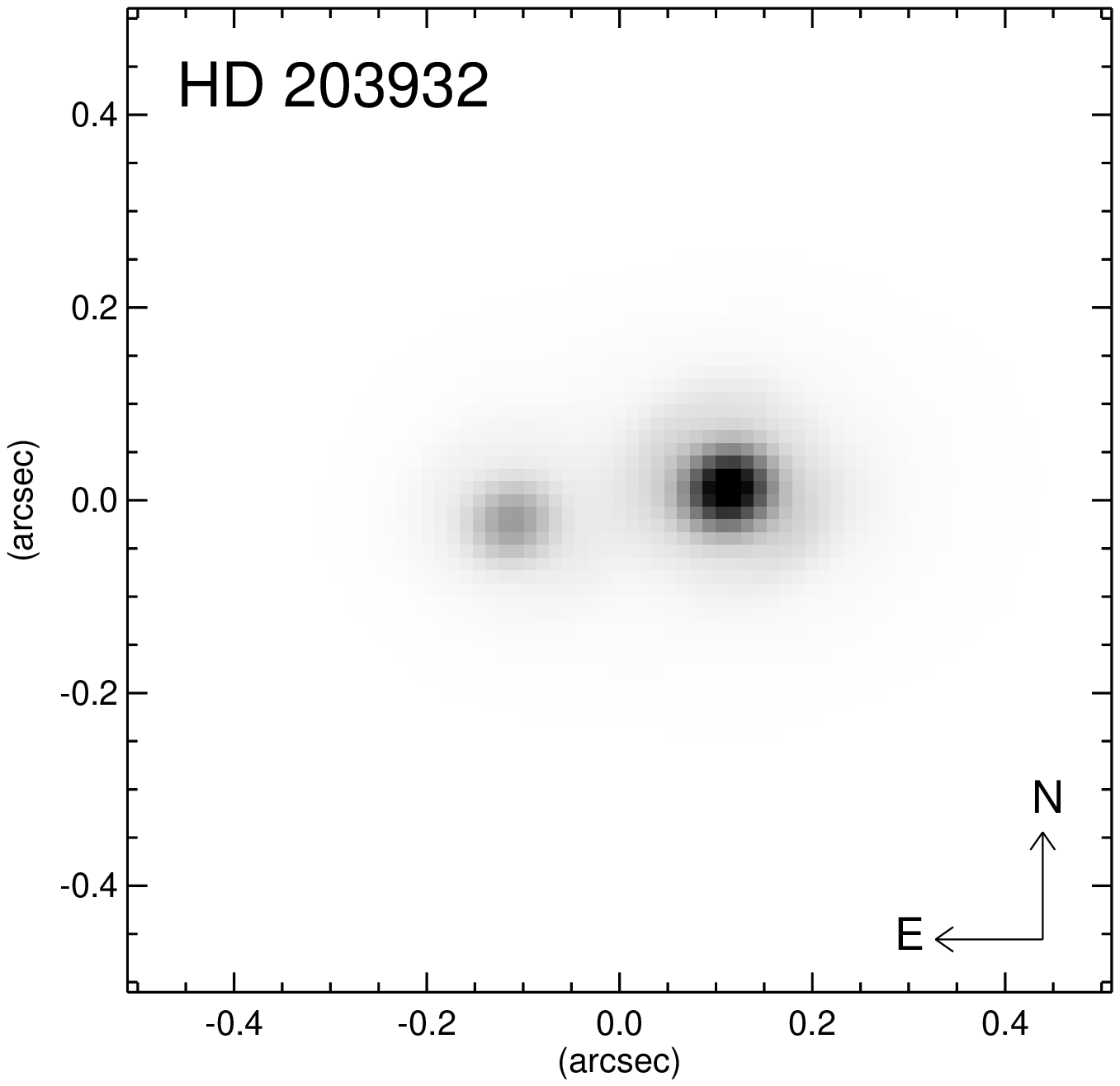}
\includegraphics[width=0.32\textwidth, angle=0]{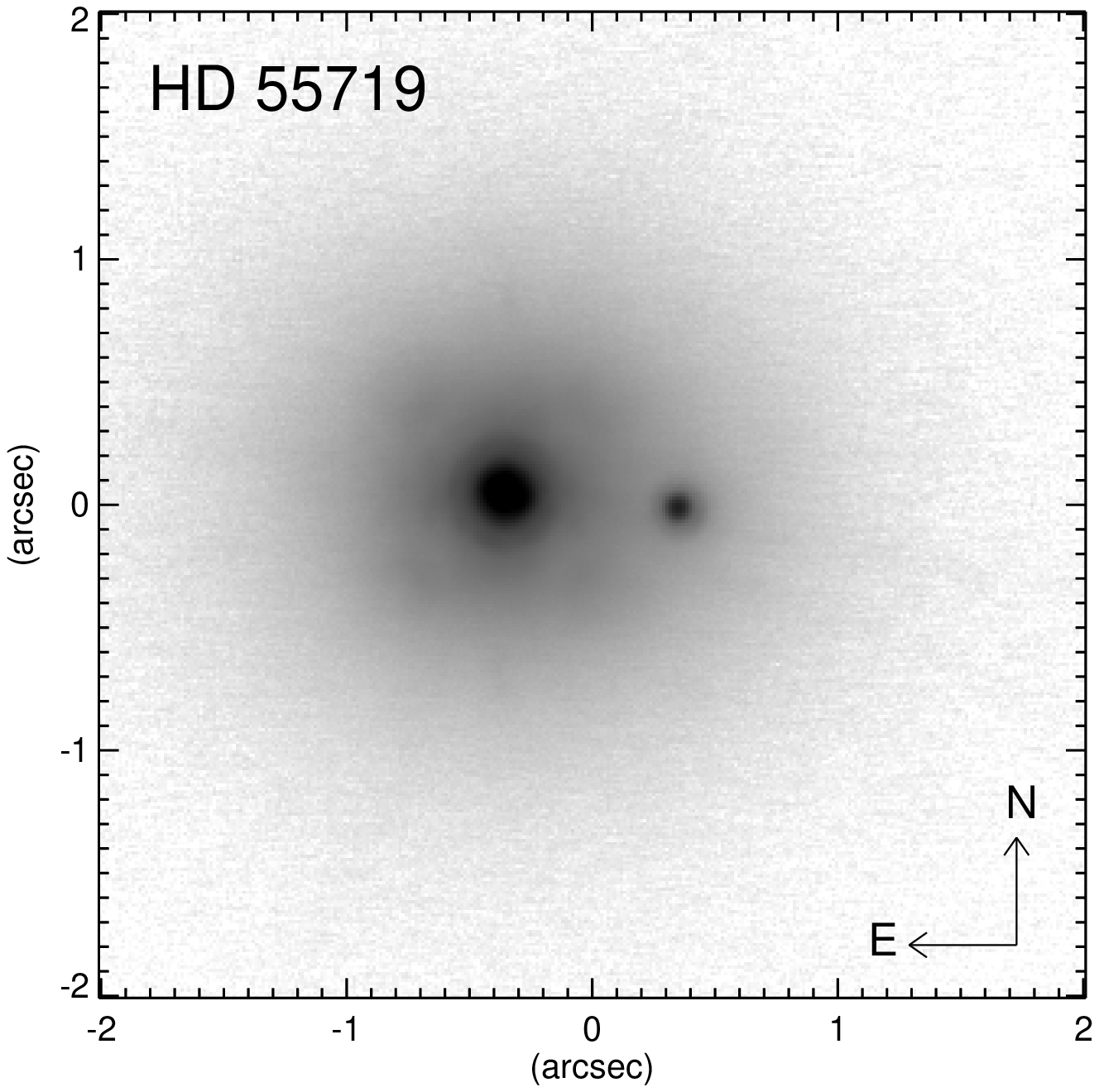}
\caption{
Images of the potential binary systems detected in our VLT/NACO survey.
}
\label{fig:binaries}
\end{figure*}

\begin{figure*}
\centering
\includegraphics[width=0.30\textwidth, angle=0]{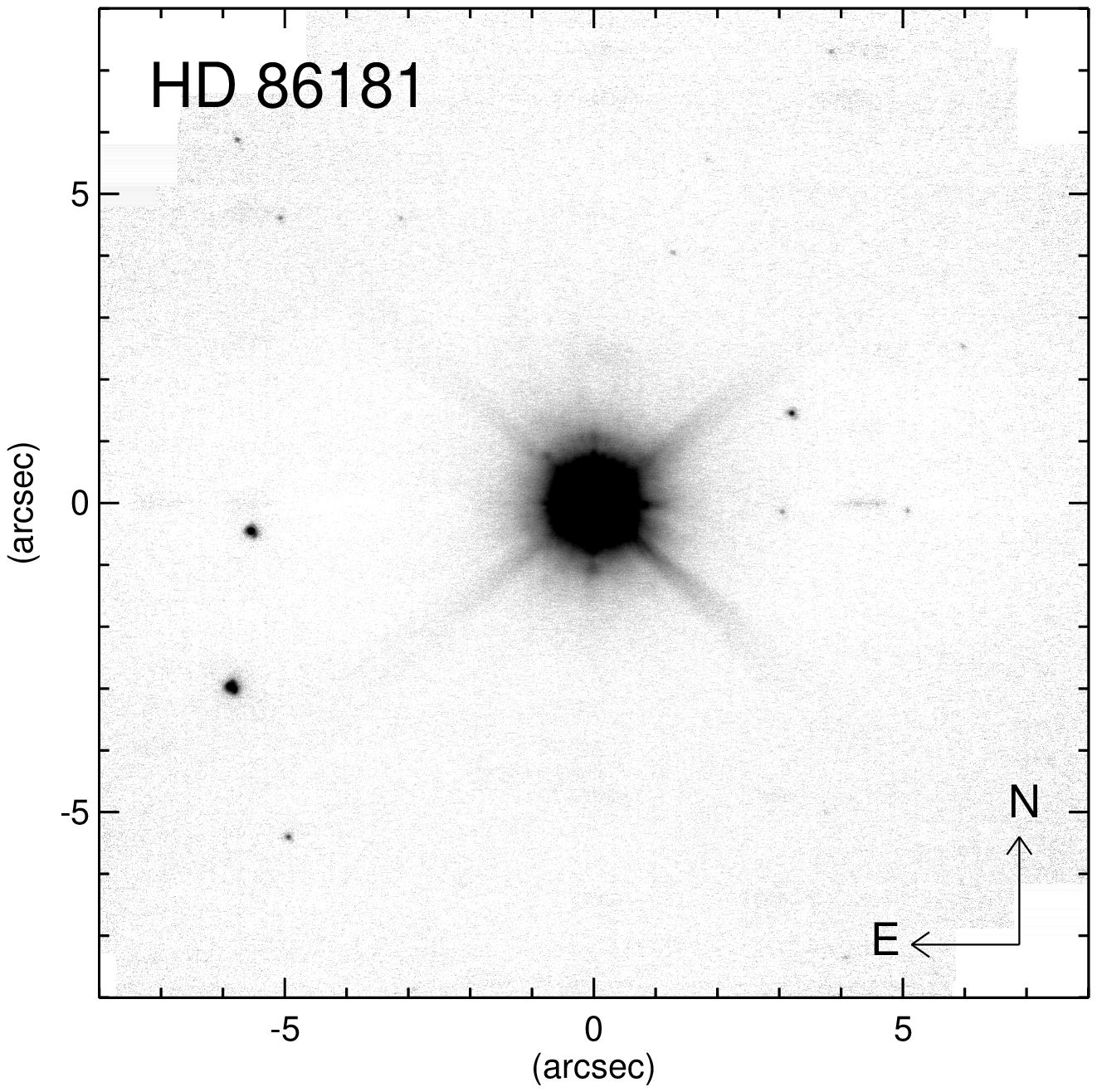}
\includegraphics[width=0.30\textwidth, angle=0]{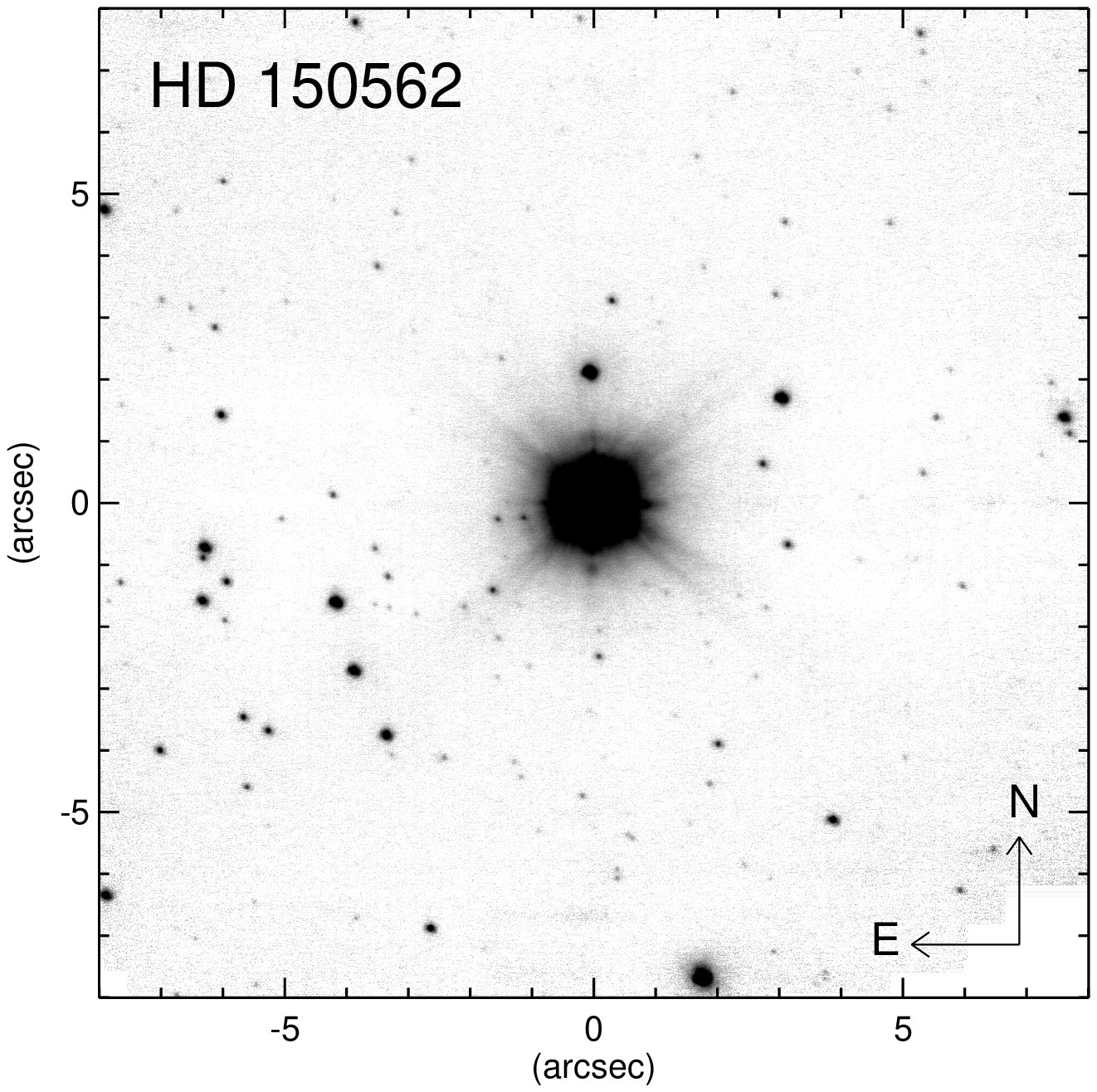}
\includegraphics[width=0.30\textwidth, angle=0]{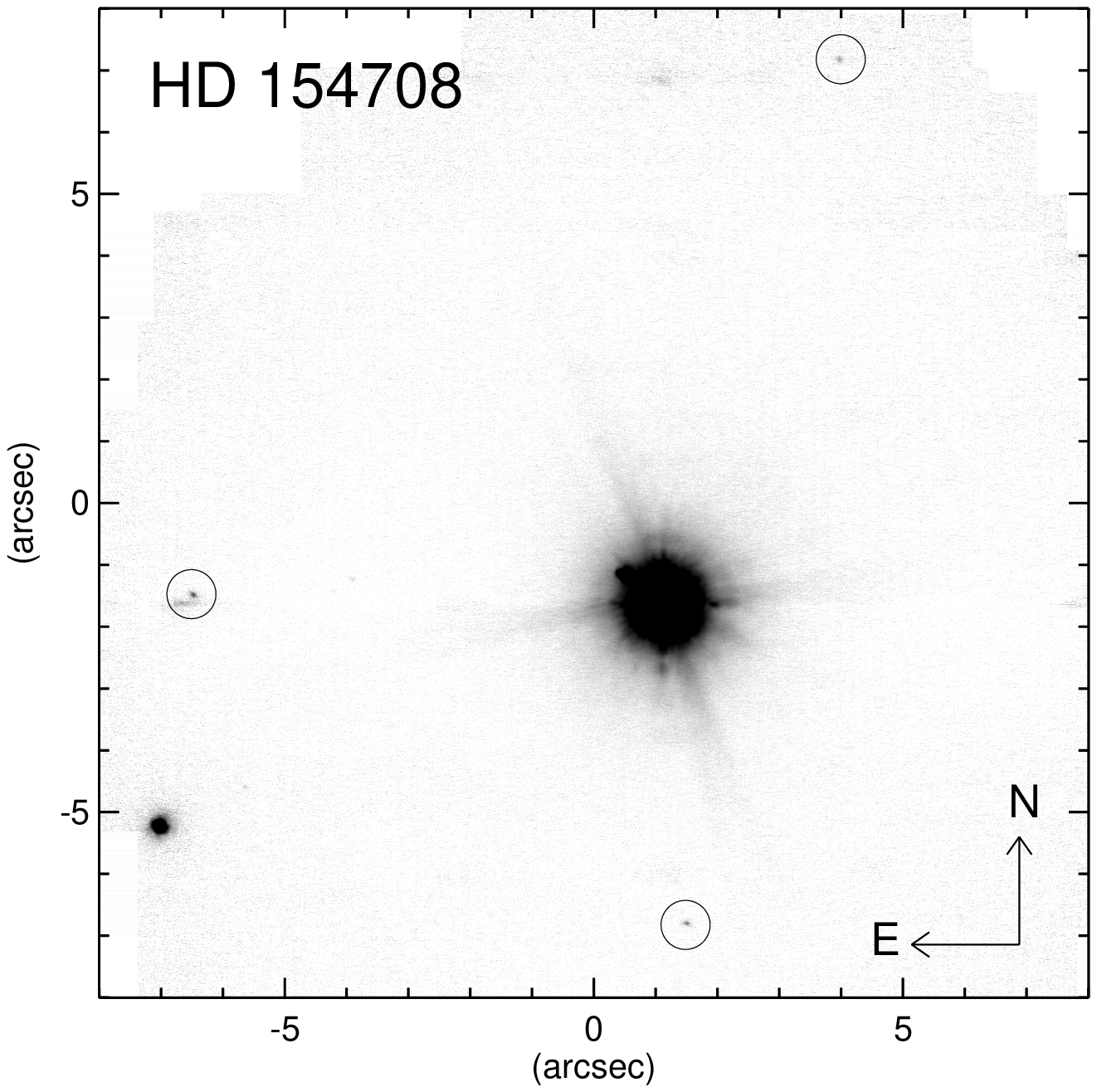}
\includegraphics[width=0.30\textwidth, angle=0]{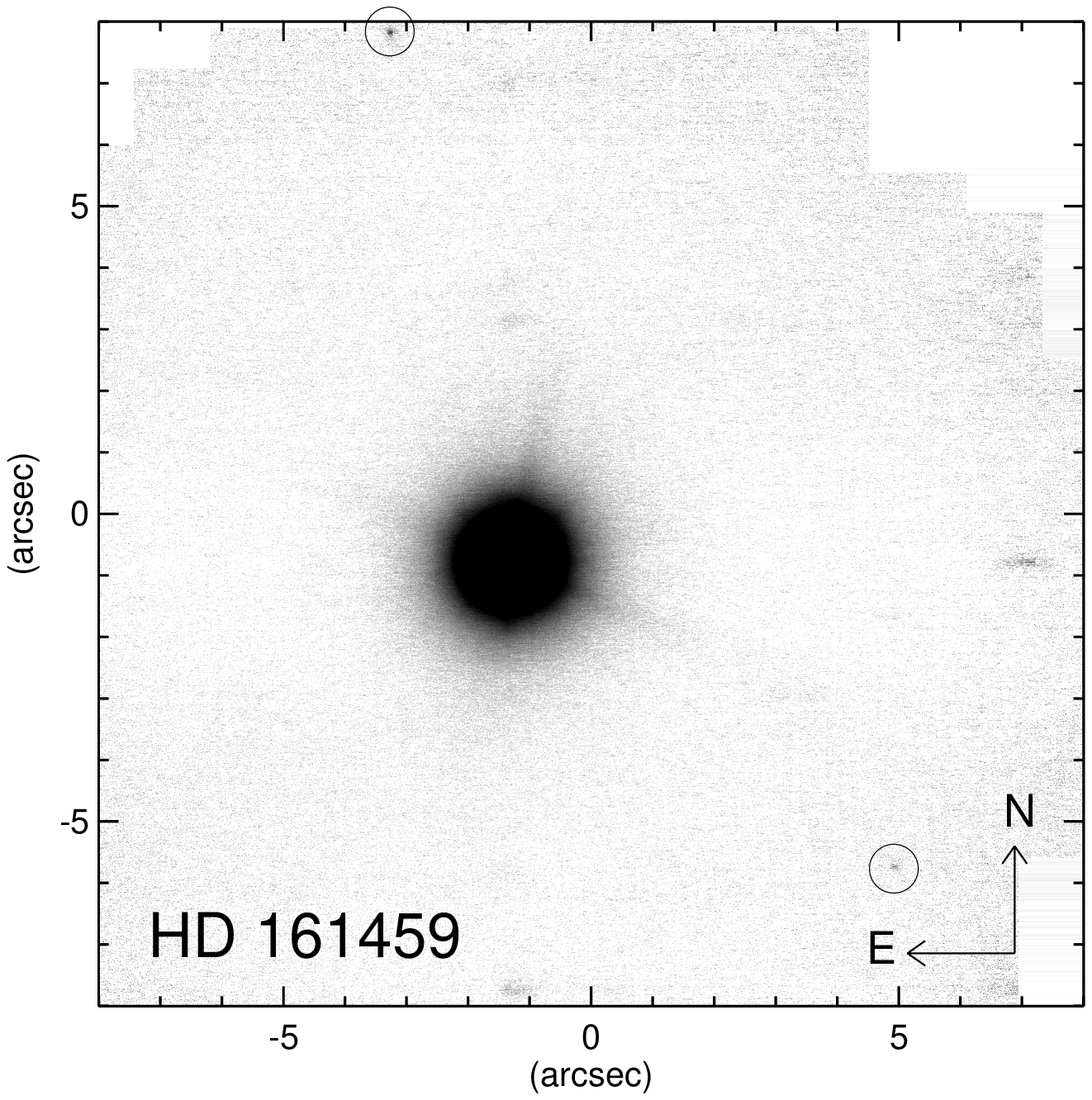}
\includegraphics[width=0.30\textwidth, angle=0]{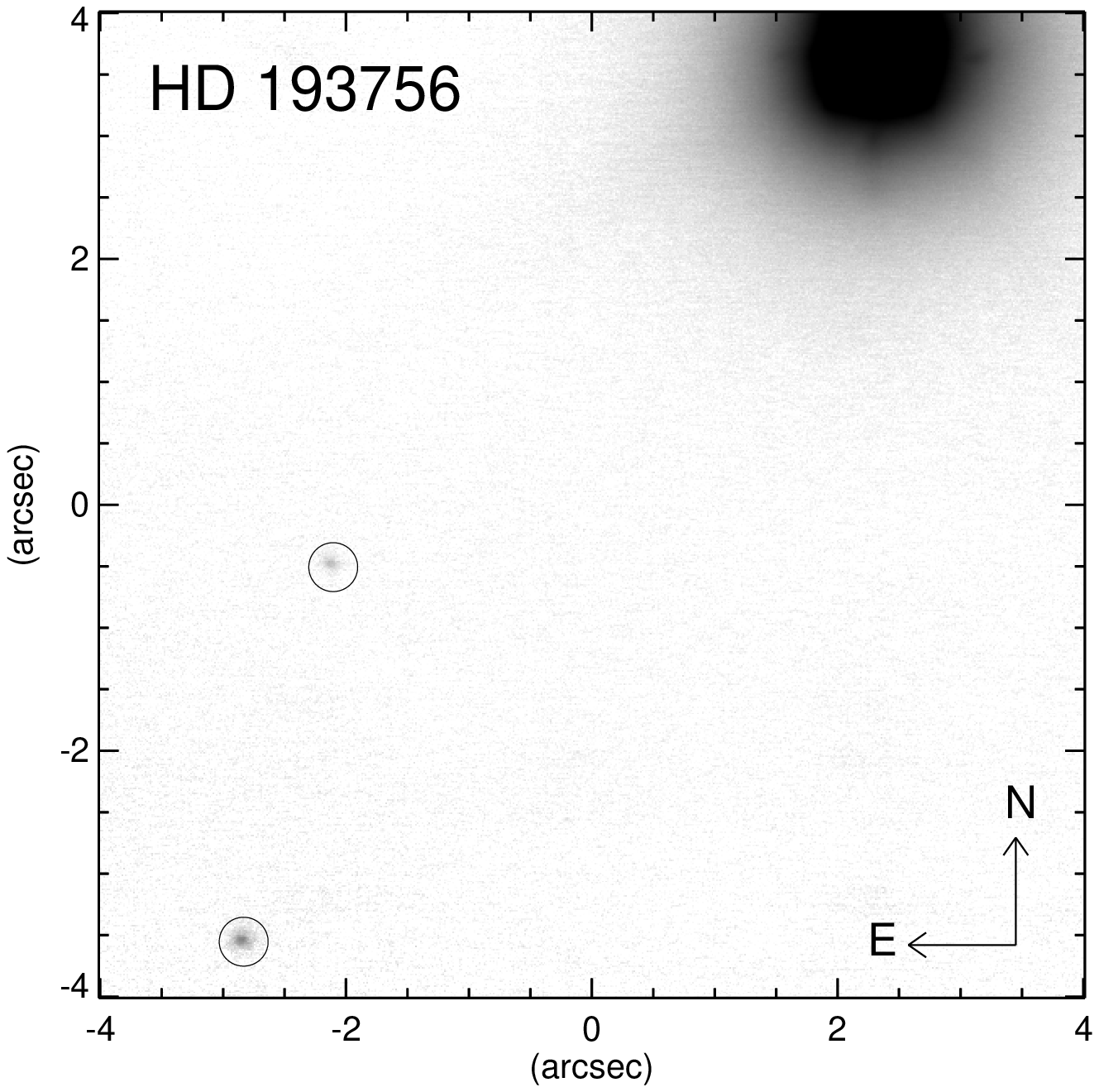}
\includegraphics[width=0.30\textwidth, angle=0]{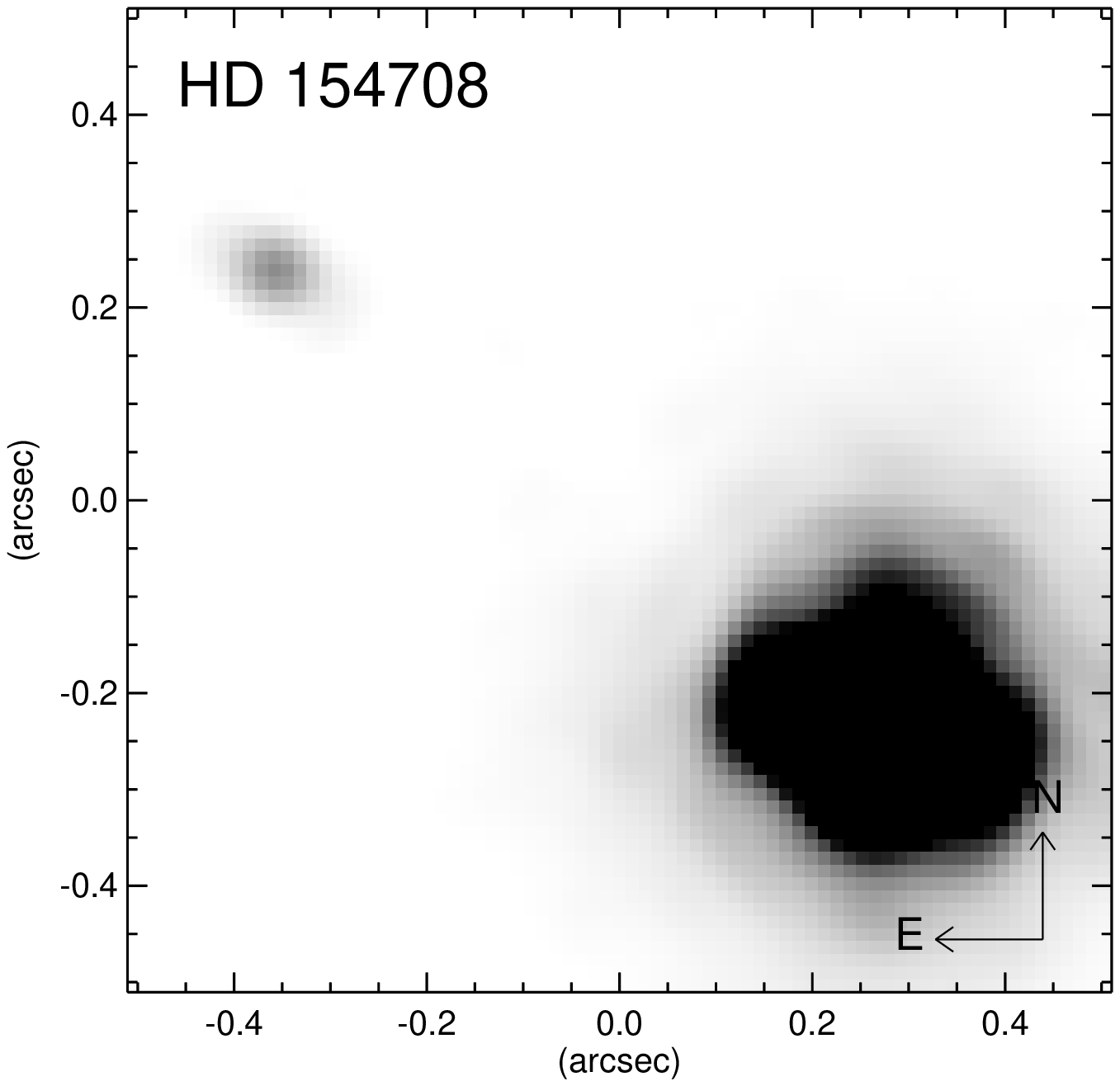}
\caption{
Images of the systems with more than one companion candidate detected in our VLT/NACO survey.
}
\label{fig:triples}
\end{figure*}

The images of the resolved potential binary systems are shown in Fig.~\ref{fig:binaries},
while the systems with more than one companion candidate can be found in Fig.~\ref{fig:triples}.
All images are displayed using a logarithmic scale.
In the images with the closest companion candidates, showing just the inner 1\arcsec{}, this logarithmic scale
had to be adapted to enhance the image details.
While we have tried to show all companion candidates in these images,
please note that this was not possible for HD\,86181 and HD\,150562.
The only companion candidate that is likely physical in the systems with more than one companion candidate
can be seen on the lower right of Fig.~\ref{fig:triples}.

\subsection{Limits for undetected companions and completeness}
\label{sect:completeness}

\begin{figure}
\centering
\includegraphics[width=0.45\textwidth, angle=0]{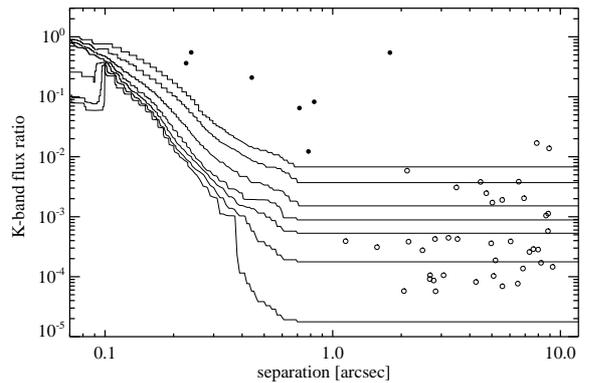}
\caption{
Completeness map for our survey.
The lines represent the completeness of our observations as derived
from the sensitivity limit for undetected companion candidates (see Sect.~\ref{sect:completeness}).
The lines are, from top to bottom, 99\%, 90\%, 70\%, 50\%, 30\%, 10\%, and 1\% completeness.
The full circles represent the companion candidates found in our study that we consider to be physical,
while the open circles represent the companion candidates we consider to be chance projections.
}
\label{fig:compl_maps}
\end{figure}

The detection limits were computed using the method described in Correia et al.\ (\cite{Correia2006}).
At each radial distance and position angle from the star, the standard deviation in the flux was 
calculated over a circular region of radius 70\,mas, i.e., equivalent 
to the mean size of the point spread function (PSF) core.
The detection limit as a function of separation from the star
is the average of the 5$\sigma$ flux over 
all position angles except those lying in the direction of the companion candidate.

For any given component of the potential binary systems as well as for the unresolved sources,
we thus have the limiting flux ratio for undetected companion candidates as a function of separation.
We can therefore produce a completeness map for 
each of these sources, i.e.\ a map giving the probability of detecting
a companion candidate as a function of separation and magnitude difference.
The total completeness map of all 
sources is simply the average of all individual completeness maps.

Figure~\ref{fig:compl_maps} shows the total completeness map.
The lines represent the completeness of our observations at levels
99\%, 90\%, 70\%, 50\%, 30\%, 10\%, and 1\% completeness,
from top to bottom.
The full circles represent the companion candidates found in our study that we consider to be physical,
i.e.\ with low chance projection probabilities (see Sect~\ref{sect:projections}),
while the open circles represent the companion candidates we consider to be chance projections.
All companion candidates that we consider physical fall above the 
99\% completeness level, while all but two companion candidates that we consider as
chance projections fall below the 99\% completeness level, and those two
are at the edge of the field.
Please note that our method is not very accurate inside of the first Airy ring.  
Overall, we are confident in the completeness levels. 
Assuming that both the distribution of separation and the 
distribution of flux ratio of companion candidates are flat (which is obviously
a very rough assumption), we estimate
that the completeness is above 90\% for separations of
a) between 0\farcs25 and 0\farcs4 and above a flux ratio of 10$^{-1}$,
b) between 0\farcs4 and 1\arcsec{} and above a flux ratio of 10$^{-2}$,
and c) between 1\arcsec{} and 8\arcsec{} and above a flux ratio of 5$\times$10$^{-3}$.
It should be noted that the S13 camera of NACO is incomplete in the detection of companion candidates 
at large separations ($>\sim$7--8\arcsec{}).



\subsection{Chance projections}
\label{sect:projections}

To identify the systems whose components are 
gravitationally bound and those that are only the result of a
chance projection, we used a statistical approach
(see e.g.\ Correia et al.\ \cite{Correia2006}).
In a first step, we determined the local surface density of background/foreground sources in each field.
For this purpose, we compiled the number of 2MASS objects brighter
than the companion candidates in the K-band in a 30\arcmin{}$\times$30\arcmin{}
field surrounding each primary.
This leads to the average surface density of objects brighter than the limiting 
magnitude $\Sigma(K < K_{\rm comp})$.
Assuming a random uniform 
distribution of unrelated objects across the field, the resulting 
probability $P(\Sigma, \Theta)$ of at least one unrelated source being located 
within a certain angular distance $\Theta$ from a particular target is 
given by 

\begin{displaymath}
P(\Sigma, \Theta) = 1 - e^{-\pi\Sigma\Theta^2}.
\end{displaymath}


The last column of Table~\ref{tab:astrometry} gives the resulting probability for a companion candidate to be unrelated
to the primary of the system.
Since the 2MASS Point Source Catalog is incomplete for stars fainter than
${\rm K}=14.3$, the calculated chance projection probabilities are only lower limits
for sources fainter than ${\rm K}=14.3$.
This is the case for all sources with high chance projection probability
except for HD\,154708AB, which is ${\rm K}=12.62$ and has a chance projection probability
of 3.88\%, while all companion candidates with a low chance projection probability
are brighter than ${\rm K}=13$.
Out of the eight companion candidates detected in our survey in potential binary systems, five 
have probabilities to be projected unrelated stars well below the percent level.
This means that they are very likely bound to their primaries, although considering probabilities of individual sources is known to be prone 
to error (see e.g.\ Brandner et al.\ \cite{Brandner2000} for a discussion).
The three other companion candidates have chance projection probabilities on the order of 6\%.
In the systems with more than one companion candidate, all except one companion candidate (HD\,154708AC) are very likely chance projections,
with chance projection probabilities between 3\% and 57\%.
On the other hand, while the vast majority of the objects with high chance projection probabilities
should be background stars, some of them could have been captured (see e.g.\
Kouwenhoven et al.\ \cite{Kouwenhoven2010}; Moeckel \& Clarke \cite{MoeckelClarke2011}).

\section{Discussion}
\label{sect:discussion}

\begin{figure}
\centering
\includegraphics[width=0.45\textwidth, angle=0]{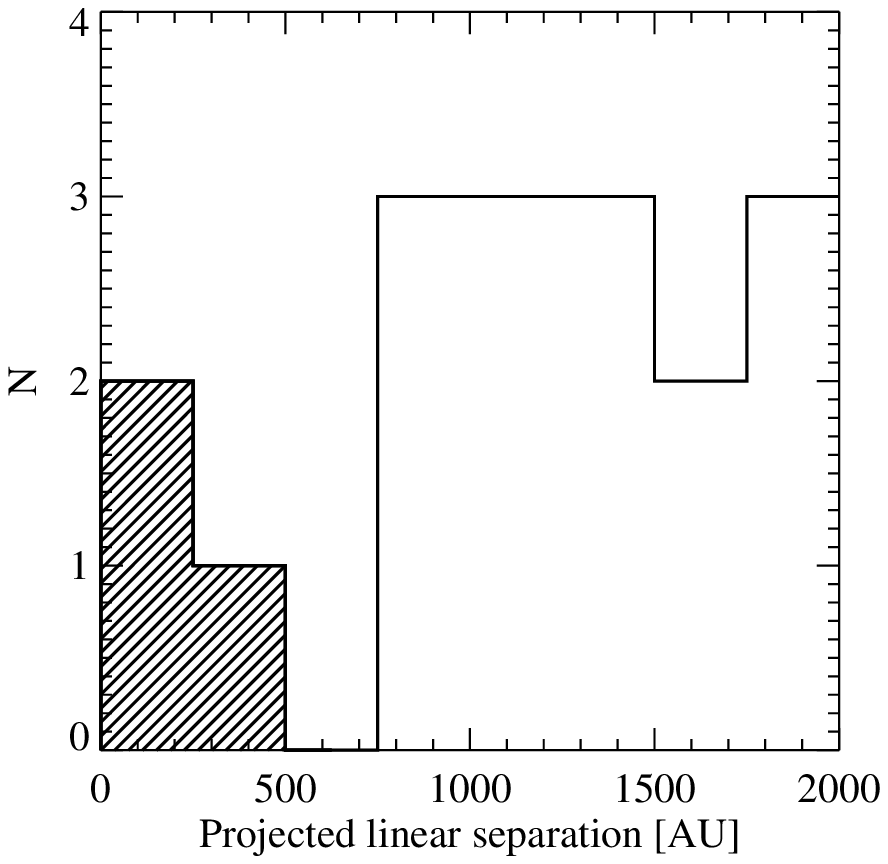}
\caption{
Distribution of the projected separations of the studied systems
with roAp primaries.
For this figure, we have used all companion candidates where a parallax was
available for the roAp star.
The shaded region shows the three objects where the chance projection
probability is smaller than 1\%.
}
\label{fig:histo}
\end{figure}

\begin{table}
\centering
\caption{
Overview of the known multiplicity of the objects studied in this article.
}
\label{tab:multi}
\begin{tabular}{rccc}
\hline
\hline
\multicolumn{1}{c}{HD} &
\multicolumn{1}{c}{SB1} &
\multicolumn{1}{c}{Astrometric} &
\multicolumn{1}{c}{Comment} \\
\multicolumn{1}{c}{} &
\multicolumn{1}{c}{} &
\multicolumn{1}{c}{or Visual} &
\multicolumn{1}{c}{} \\
\hline
6532   &    &   & \\
9289   &    & 1 & ND \\
12932  &    & 1 & ND \\
19918  &    &   & \\
24712  &    &   & \\
42659  &    &   & \\
86181  &    &   & \\
99563  &    & 1 & CD \\
101065 &    &   & \\
116114 & X? & 1 & NR \\
119027 &    &   & \\
122970 &    &   & \\
134214 &    &   & \\
137949 &    &   & \\
150562 &    &   & \\
154708 &    & 1 & ND \\
161459 &    &   & \\
166473 &    &   & \\
176232 &    &   & \\
185256 &    &   & \\
190290 &    &   & \\
193756 &    &   & \\
196470 &    &   & \\
201601 &    & 3 & CD+NR+NR\\
203932 &    & 1 & ND \\
213637 &    &   & \\
217522 &    &   & \\
218495 &    &   & \\
\hline
\multicolumn{3}{c}{Other A stars} \\
\hline
40711  & X? &   & \\
55719  & X  & 1 & CD \\
59435  & X? &   & \\
\hline
\end{tabular}  
\tablefoot{
An X in the second column denotes an SB1 system; the ? indicates objects
where there are only hints for an SB1 system.
In the last column we point out the new detections (ND), the confirmed
detections (CD), and the not reachable companions (NR).
The not reachable companions are either potentially very narrow (HD\,116114)
or outside of our field-of-view (HD\,201601).
}
\end{table}

Here, we announce the detection of 45 companion candidates in 13 of the observed 28 roAp stars,
with 39 of these very likely being chance projections.
They have K magnitudes between 6\fm8 and 19\fm5 and angular
separations ranging from 0\farcs23 to 8\farcs9,
corresponding to linear projected separations of 30--2400\,AU.
For the eight companion candidates in  potential binary systems, three have a high probability to be chance projections.
In the systems with more than one companion candidate, all companion candidates but HD\,154708AC are very likely chance projections.
All companion candidates, except for the companion candidates of HD\,99563 and HD\,201601,
were detected by us for the first time.
The companion candidate for the noAp star HD\,55719 was also known before.
In Fig.~\ref{fig:histo} we show the distribution of the projected linear separations for the studied
multiple systems with roAp primaries.
The distribution is missing the 26 companion candidates where no parallax information is available
for the roAp star.
For only three companion candidates with low probability to be a chance projection exist
parallax data.
These are the three objects with a projected linear separation below 500\,AU.

In our survey, we found in six of the 28 studied systems with an roAp primary
one visual companion candidate that has a high probability to be a physical
companion, resulting in a multiplicity fraction of $21\pm9$\%.
This is low compared with similar surveys of A type stars.
On the other hand, should all of the objects we believe to be chance projections turn out
to be physical companions, the multiplicity fraction may be as high as $46\pm13$\%.
We introduced a bias in the sample by removing the well studied $\alpha$\,Cir, which
has a companion.
Inserting $\alpha$\,Cir in the sample, we get a multiplicity fraction of $24\pm9$\%.

Kouwenhoven et al.\ (\cite{Kouwenhoven2005}) studied the binarity of A and B stars in the
OB association Sco~OB2 with adaptive optics using a $Ks$ filter and a similar field-of-view.
65 of the 199 stars in their sample have at least one companion candidate, leading to a binary fraction
of $33\pm4$\%.
If one restricts the survey to the 113 A type stars, there are 40 stars showing multiplicity,
giving a multiplicity fraction of $35\pm6$\%.
We should note however that the distance to the stars in our sample is on average higher compared
with the 130\,pc to Sco~OB2, with nearly half of the stars in our sample having no parallax determined.
Ehrenreich et al.\ (\cite{Ehrenreich2010}) find in their volume limited sample of 38 late B to F-type stars,
observed with NACO and PUEO, companion candidates to 17 objects.
If one restricts their sample to the 19 A-type stars, this leaves companion candidates to seven objects, or a binary
fraction of $37\pm14$\%.
While Ehrenreich et al.\ employ on some of their observations a larger field-of-view,
they consider all companion candidates to A-type stars with separations larger than 7\arcsec{}
as background stars.
The distances to these A stars are lower than 67\,pc, and only two of our own sources would fall
into this distance range.
Schr\"oder \& Schmitt (\cite{ SchroederSchmitt2007}) looked at the A-type stars listed
in the Bright Star Catalog (Hoff\/leit \& Jaschek \cite{HoffleitJaschek1991})
and found for the 1966 objects listed therein a binary frequency of $16\pm1$\%.
Looking at a volume limited complete sample of all A stars up to a distance of 50\,pc,
they find 82 binaries among 220 stars, a binary frequency of $37\pm4$\%.
They identified companion candidates from catalogs, variations in radial velocity or proper motions,
variations in ROSAT X-ray light curves at different time scales, and rotational velocity.
Since Schr\"oder \& Schmitt did not make use of high spatial resolution observations, they
could have missed
a number of companion candidates, leading to an underestimation of the true object multiplicity.
Assuming that the three different surveys listed above can be summed up, this
would lead to 129 binaries for 352 systems, or a binary frequency of $37\pm3$\%.

Even taking into account 
the different observing strategies and object distances used in the studies
discussed above, we believe that the 1$\sigma$ difference
hints that our sample shows in comparison
a somewhat lower number of stars harboring a companion, assuming that all objects found to have
high chance projection probabilities are indeed no physical companions.

Since our sample is quite heterogeneous with respect to distance, we
could suffer from a bias that allows us to detect companion candidates only around the nearby
roAp stars.
We believe that this is not the case.
If we separate the sample into three groups with different parallaxes,
then we find one companion candidate with low chance projection probability among the five objects
with parallaxes above 10\,mas, two among the ten objects with parallaxes 
below 10\,mas, and three among the 13 objects with no parallax measurement,
not favoring nearby systems.

In Table~\ref{tab:multi}, we present the list of the observed
roAp stars with notes about their multiplicity.
For each object, we indicate whether it is known to be an SB1
and how many astrometric or visual high probability companion candidates are known.
Note that there are no known SB2 in our sample.
Since the roAp stars have been observed quite extensively for
radial velocity variations, the probability that spectroscopic
binaries have been missed is quite low.
On the other hand, it is not impossible that long term radial
velocity variations have been overlooked for individual sources.
Of the 28 roAp stars studied, only HD\,116114 is a potential SB1 system.
It is also an astrometric binary.
Six objects have visual high probability companion candidates.
Another seven stars have visual companion candidates that are very likely chance
projections.
15 roAp stars do not show any hint at being part of a binary.
We also looked into why the found companion candidates were not
detected in radial velocity surveys.
For the companion of HD\,201601, which is the closest we found
in our sample with a projected linear separation of 30.1\,AU, we used the orbital parameters
published by Stelzer et al.\ (\cite{Stelzer2011}) and computed a radial
velocity amplitude of 1.7\,km\,s$^{-1}$ for the primary.
Kochukhov \& Ryabchikova (\cite{KochukhovRyabchikova2001}) found the
largest pulsation amplitudes of 0.8\,km\,s$^{-1}$ for spectral lines of
rare-earth elements.
We believe that the accuracy and time span of radial velocity measurements obtained
so far is not sufficient to detect radial velocity changes induced by the companion
for HD\,201601.
The other likely companions have projected linear separations of 455.1\,AU (HD\,99563),
968.4\,AU (HD\,101065), and 115.9\,AU (HD\,154708AC), which lead to even lower
radial velocity amplitudes.

Our recent study (Sch\"oller et al.\ \cite{Schoeller2010}) of another group of chemically peculiar stars with Hg and Mn 
overabundances using diffraction-limited near-infrared imaging with NACO led to 
the detection of 34 near IR companion candidates for the
57 stars studied, confirming that this type of chemically peculiar stars 
is frequently formed in multiple systems.
The interpretation of the difference in duplicity between roAp and HgMn stars and its meaning
for the understanding of the origin of the chemical anomalies and pulsations in roAp stars is not straightforward.
We assume that most late B-type stars formed in binary systems with certain orbital parameters become HgMn stars
(e.g., Hubrig \& Mathys \cite{HubrigMathys1995}; Hubrig et al.\ \cite{Hubrig2007}; Gonz{\'a}lez et al.\ \cite{Gonzalez2010};
Hubrig et al.\ \cite{Hubrig2006}).
Our results hint at the possibility that
magnetic Ap stars become roAp stars if they are not born in a close binary system.

Tidal forces might conceivably also play a non-negligible role in systems with a larger 
separation, provided that their eccentricity is large enough.
Interaction would then occur mostly on the part of the orbit when the components are closest, since tidal 
forces are strongly dependent on the distance between the components.
At present, though, almost nothing is known about the orbital eccentricities of the noAp binaries.
For the roAp star HD\,201601, Stelzer et al.\ (\cite{Stelzer2011}) give an eccentricity of 0.56$\pm$0.05.


On general grounds, the issue of whether duplicity affects pulsation through tidal 
interaction is unsettled. From the theoretical point of view, while some authors 
(e.g., Cowling \cite{Cowling1941}; Zahn \cite{Zahn1977}) have conjectured that 
tides in close binary systems 
may act as an external perturbing force driving oscillations, the question whether 
tidal interaction may also be efficient in damping already existing pulsations does 
not seem to have ever been addressed.
Recently, significant advances have been made in tidally-driven pulsations in eccentric binaries 
with Kepler data.
There are 18 heartbeat stars in Thompson et al.\ (\cite{Thompson2012}), one of which is KOI-54
(Welsh et al.\ \cite{Welsh2011}; Burkart et al.\ \cite{Burkart2012}; Fuller \& Lai \cite{FullerLai2012}).
The theory of Kumar et al.\ (\cite{Kumar1995}) fits these heartbeat stars beautifully.
Observationally, in the same region of the 
parameter space in which pulsations were detected, there is only one binary system 
with a noAp primary presently known, in which the two components are close enough 
so that significant tidal interaction occurs between them
(Giuricin et al.\ \cite{Giuricin1984}), the SB1 HD\,200405 with $P$ = 1.63\,d (North \cite{North1998}).

The results of our study support our suspicion
that roAp stars are rarely found in binary and multiple systems.
However, companionship can not be established based on K photometry alone, and
confirming the nature with a near-infrared spectrograph is essential for
establishing their true companionship.
Future spectroscopic observations in the near-infrared should be used
to determine the mass of the companions much more accurately,
and explore the physics in their atmospheres by comparison of observed and
synthetic spectra.


\appendix

\section{Notes on individual systems}
\label{sect:individual}

%
%
%
%
%

\subsection{Systems unresolved in our study}
\label{sect:unresolved}

%
%
%


{\it HD\,116114:}
This star was marked by Renson \& Manfroid (\cite{RensonManfroid2009})
as a potential binary with a period of 4000\,d.
Dommanget \& Nys (\cite{DommangetNys2002}) mark this star in the CCDM catalog
as an astrometric binary from {\em Hipparcos}.

%
%
%
%
%
%
%
%
%


{\it HD\,40711:}
This star was marked by Renson \& Manfroid (\cite{RensonManfroid2009})
as a potential binary with a period of 1245\,d.

{\it HD\,59435:}
This star was marked by Renson \& Manfroid (\cite{RensonManfroid2009})
as a potential binary with a period of 1386\,d.

{\it HD\,6532, HD\,19918, HD\,24712, HD\,42659,
HD\,119027, HD\,122970, HD\,134214, HD\,137949, HD\,166473, HD\,176232, HD\,190290, HD\,213637, HD\,217522, and HD\,218495:}
There are no references in the literature that indicate multiplicity for these objects.

\subsection{Systems resolved in our study, but very likely chance projections}
\label{sect:binaries_chance}

{\it HD\,86181:}
There are no references in the literature that indicate multiplicity for this object.
We find a total of eleven objects between K magnitudes 15 and 19, with separations
of 3\arcsec{} to 8\arcsec{} around this star.
All of them have chance projection probabilities above 10\%.

{\it HD\,101065:}
There are no references in the literature that indicate multiplicity for this object.
We detect a new companion candidate to this star
at a separation of 8\farcs648 and a position angle of 140.9$^{\circ}$.
A chance projection probability above 5\% suggests that this companion candidate might not be a
physical companion.

{\it HD\,150562:}
There are no references in the literature that indicate multiplicity for this object.
We find a total of 17 objects between K magnitudes 14.5 and 19.5, with separations
of 1\farcs1 to 8\arcsec{} around this star.
All of them have chance projection probabilities above 3\%.

{\it HD\,161459:}
There are no references in the literature that indicate multiplicity for this object.
We find two objects with K magnitudes around 18, with separations
of 8\arcsec{} and 9\arcsec{} around this star.
Both have chance projection probabilities above 40\%.

{\it HD\,185256:}
There are no references in the literature that indicate multiplicity for this object.
We detect a new companion candidate to this star
at a separation of 4\farcs965 and a position angle of 306.7$^{\circ}$.
A chance projection probability above 6\% suggests that this companion candidate might not be a
physical companion.

{\it HD\,193756:}
There are no references in the literature that indicate multiplicity for this object.
We find two objects with K magnitudes around 16 and 17, with separations
of 9\arcsec{} and 6\arcsec{} around this star.
They have chance projection probabilities of 8\% and 3\%.

{\it HD\,196470:}
There are no references in the literature that indicate multiplicity for this object.
We detect a new companion candidate to this star
at a separation of 6\farcs921 and a position angle of 190.9$^{\circ}$.
A chance projection probability above 6\% suggests that this companion candidate might not be a
physical companion.

\subsection{Systems resolved in our study}
\label{sect:binaries_real}

{\it HD\,9289:}
There are no references in the literature that indicate multiplicity for this object.
We detect a new companion candidate to this star
at a separation of 0\farcs441 and a position angle of 72.7$^{\circ}$.

{\it HD\,12932:}
There are no references in the literature that indicate multiplicity for this object.
We detect a new companion candidate to this star
at a separation of 0\farcs239 and a position angle of 171.4$^{\circ}$.

{\it HD\,99563:}
The Washington Double Star Catalog (Mason et al.\ \cite{Mason2001}) lists a companion at a separation of 1\farcs{}8
and a position angle of 218$^{\circ}$.
Dommanget \& Nys (\cite{DommangetNys2002}) list this component in the CCDM catalog
at a separation of 1\farcs7 and a position angle of 213$^{\circ}$.
We find this companion at a separation of 1\farcs{}784 and a position angle of 216.6$^{\circ}$.

{\it HD\,201601:}
The Washington Double Star Catalog (Mason et al.\ \cite{Mason2001}) lists one companion at a distance between
1\farcs5 and 2\farcs1 and position angles between 264$^{\circ}$ and 277$^{\circ}$,
a second companion at a distance between 25\farcs0 and 57\farcs3, and a third
companion at a distance of about 6\arcmin{}.
We find the close companion at a distance of 0\farcs829 and a position angle of 256.8$^{\circ}$.
Stelzer et al.\ (\cite{Stelzer2011}) used this new measurement and combined it with the data from
the Washington Double Star Catalog and from {\em Hipparcos} to derive a preliminary orbit for $\gamma$\,Equ
with a period of 274.5\,yr.
They estimated the mass of the companion to be 0.6$\pm$0.4\,$M_{\sun}$.

{\it HD\,203932:}
There are no references in the literature that indicate multiplicity for this object.
We detect a new companion candidate to this star
at a separation of 0\farcs227 and a position angle of 98.2$^{\circ}$.

{\it HD\,154708:}
There are no references in the literature that indicate multiplicity for this object.
We find five objects around this star.
Only the component AC, with a separation of 0\farcs782 at position angle 53.0$^{\circ}$ 
and a K magnitude of 12.75 has a chance projection probability below $10^{-3}$.
The other four objects between K magnitudes 12.5 and 17.5, with separations
of 5\arcsec to 9\arcsec{} have chance projection probabilities above 3\%.

{\it HD\,55719:}
This system is an SB1 with a period of 46.3140\,d,
according to the 9$^{\rm th}$ Catalog of Spectroscopic Binary Orbits (Pourbaix et al.\ \cite{Pourbaix2004}).
The Washington Double Star Catalog (Mason et al.\ \cite{Mason2001}) lists a companion at a distance of 0\farcs8
and a position angle of 258$^{\circ}$.
We find the companion at a separation of 0\farcs714 and a position angle of 265.0$^{\circ}$.

\begin{acknowledgements}


We would like to thank Christian Hummel for discussions on the orbit
of HD\,201601.
Part of this work was supported by the ESO Director General
Discretionary Fund.
This publication made use of data products from the Two Micron All
Sky Survey and the DENIS database,
and of the SIMBAD database, operated at CDS, Strasbourg, France.

\end{acknowledgements}

\end{document}